\documentclass[journal,twoside,web]{ieeecolor}
\usepackage{tmi}
\usepackage{cite}
\usepackage{amsmath,amssymb,amsfonts}
\usepackage{algorithmic}
\usepackage{graphicx}
\usepackage{textcomp}
\usepackage{hyperref}
\usepackage{soul}
\usepackage{colortbl}

\def\BibTeX{{\rm B\kern-.05em{\sc i\kern-.025em b}\kern-.08em
    T\kern-.1667em\lower.7ex\hbox{E}\kern-.125emX}}
\begin{document}
\title{Knee Osteoarthritis Severity Prediction using an Attentive Multi-Scale Deep Convolutional Neural Network}
\author{Rohit Kumar Jain,
        Prasen Kumar Sharma,
        Sibaji Gaj,
        Arijit Sur
        and Palash Ghosh
        \thanks{This work has been submitted to the IEEE for possible publication. Copyright may be transferred without notice, after which this version may no longer be accessible.}
\thanks{Rohit kumar Jain, Prasen Kumar Sharma, and Arijit Sur are with Department of Computer Science and Engineering, Indian Institute of Technology Guwahati, India.} \thanks{Sibaji Gaj is with Cleveland Clinic, Ohio, USA.} \thanks{Palash Ghosh is with Department of Mathematics, Indian Institute of Technology Guwahati, India, and Centre for Quantitative Medicine, Duke-NUS Medical School, National University of Singapore, Singapore.}
\thanks{Corresponding author: Rohit Kumar Jain}
\thanks{Email: \texttt{jkrohit03@gmail.com}}}

\maketitle

\begin{abstract}
Knee Osteoarthritis (OA) is a destructive joint disease identified by joint stiffness, pain, and functional disability concerning millions of lives across the globe. It is generally assessed by evaluating physical symptoms, medical history, and other joint screening tests like radiographs, Magnetic Resonance Imaging (MRI), and Computed Tomography (CT) scans. Unfortunately, the conventional methods are very subjective, which forms a barrier in detecting the disease progression at an early stage. This paper presents a deep learning-based framework, namely {OsteoHRNet}, that automatically assesses the Knee OA severity in terms of Kellgren and Lawrence (KL) grade classification from X-rays. 
As a primary novelty, the proposed approach is built upon one of the most recent deep models, called the High-Resolution Network ({HRNet}), to capture the multi-scale features of knee X-rays.
In addition, we have also incorporated an attention mechanism to filter out the counterproductive features and boost the performance further. 
Our proposed model has achieved the best multi-class accuracy of 71.74\% and MAE of 0.311 on the baseline cohort of the OAI dataset, which is a remarkable gain over the existing best-published works. We have also employed the Gradient-based Class Activation Maps (Grad-CAMs) visualization to justify the proposed network learning.
\end{abstract}


\section{Introduction}
\label{Introduction}
\IEEEPARstart{K}{nee} osteoarthritis is a common joint disorder caused by the eroding of the articular cartilage between the joints, which leaves the bones of the knee touching and rubbing against each other. In general, it occurs in the synovial joints and results from a combination of genetic factors, injury, and overuse \cite{OKA20081300}. Obesity, specific occupation, stress, trauma, age, gender, and family history are some well-defined risk factors \cite{Lespasio}. The pain and stiffness in the joints begin to worsen by the rigorous activity and stress compared to other inflammatory arthritis where activity and exercising improve symptoms. It can also lead to instability, joint deformity, and reduction in joint functionality \cite{Lespasio}. In addition, the distance between the knee joint begins to flatten out due to the loss of the cartilage, leading to the progression of knee OA \cite{OKA20081300}. The following key changes, described by the word \texttt{LOSS}, marks the progression of knee OA: 

\begin{itemize}
    \item \texttt{L-} ``loss of joint space'', caused by the cartilage loss,
    \item \texttt{O-} ``osteophytes formations'', projections that form along the margins of the joint, 
    \item \texttt{S-} ``subarticular sclerosi'', increase in bone density along the joint line, and 
    \item \texttt{S-} ``subchondral cysts'', caused due to holes in the bone filled with fluid along the joints \cite{LOSS}.
\end{itemize}
 
\begin{figure}[t]
  \centering
      \includegraphics[width = 0.48\textwidth]{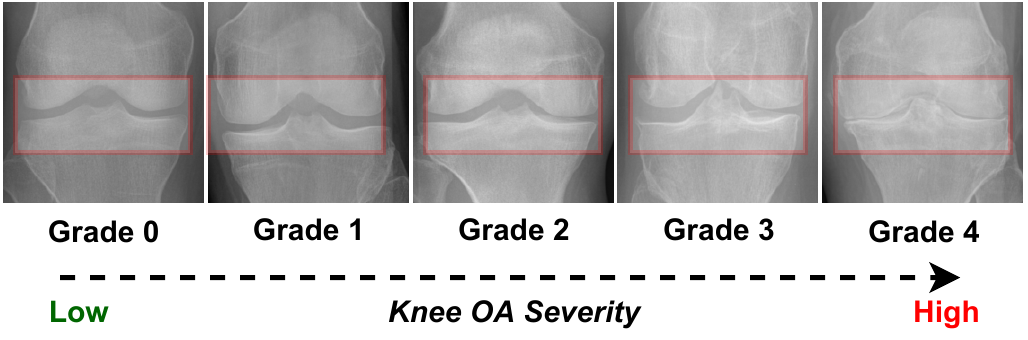}
    \caption{ \textbf{Knee OA disease progression: }A qualitative demonstration of sample X-rays and their corresponding KL grades.}
    \label{fig:Sample}
\end{figure}

Radiographic screening (X-Rays), MRI, and CT scans are a few of the common ways to detect the structural changes in the joint and diagnose knee OA's biological condition. However, the traditional treatment for knee OA may not be effective enough to completely fix the disease in today's time. Therefore, it is of utmost importance to detect the deformation of the joint at such a stage before which it becomes impossible to reverse the loss \cite{cure}. Generally, the knee OA severity is measured in terms of the World Health Organization (WHO) approved KL grading scale \cite{Kellgren494}. KL grading is a 5-point semi-quantitative progressive ordinal scale ranging from grade 0 (\textit{low severity}) to 4 (\textit{high severity}). Fig. \ref{fig:Sample} shows the disease progression along with its corresponding KL grade.  
\subsection{Challenges}
In general, a complete cure for this disease remains quite challenging to find, and OA management is mainly palliative \cite{OKA20081300,SHAMIR20091307}. MRI screenings and CT scans are effective as they highlight the three-dimensional structure of the knee joints \cite{PETERFY20081433} \cite{tmi1}. However, they have certain drawbacks, including limited availability, extreme device expenses, the time required in diagnosing, and the inclination to image ancient rarities \cite{diagnostics10080518,van}. At the same time, X-Rays are the most effective and economically feasible way of diagnosing the disease, given the routine knee OA diagnosis. However, the currently adopted methods for assessing the disease progression from X-Ray images may not be much effective. They, in general, require a very skilled practitioner to analyze the radiographic scans accurately and are thus absolutely subjective. In most cases, the practitioners require multiple tests to quantify the condition accurately, which is generally time-consuming. The analysis may differ based on their expertise and sometimes may be inaccurate. Further, multiple tests may be costly for some of the patients.

A better and in-depth understanding of knee OA may result in timely prevention and treatment. It is believed that early treatments and preventive measures are the most effective way of managing knee OA. Unfortunately, there has been no significant and predominant way of identifying the disease at an early stage to date. Recently, the use of Machine Learning (ML) and Deep Convolutional Neural Networks (CNNs) for knee OA analysis have shown remarkable supremacy in detecting even the slightest differences in biological joint structural variations in the X-Rays \cite{Kokkotis}.

Deep CNNs have been widely adopted in many medical imaging tasks, including classifications of COVID-19, pneumonia, tumor, bone fracture, polyps detection, etc. For \textit{e.g.,} CheXNet \cite{ChexNet}, a 121-layers deep CNN, performed astonishingly better than the average performance of four specialists in assessing pneumonia using plain radiographs \cite{Yadav}. However, it is difficult to collect the medical images, as the collection and annotation of such data are challenged by the expert availability, and the data privacy concerns \cite{Yadav}. 
\subsection{The Osteoarthritis Initiative (OAI) Dataset}
OAI is a distributed, observational study of patients, which is publicly available \footnote{Dataset source: \url{https://nda.nih.gov/oai/}}. It facilitates the scientific and research community worldwide to work on knee OA progression and develop new treatments and techniques beneficial for its detection and treatment. In this work, we have utilized the data acquired from the OAI repository and made available by Chen \textit{et al.} \cite{dataset, chen2019fully}. The dataset comprises knee bilateral posterior-anterior fixed flexion radiographs of 4796 participants, including male and female subjects from the baseline cohort. Fig. \ref{fig:Sample} shows sample X-ray images pertaining to each KL grade.

\section{Related Developments}
Several schemes have been developed for the Knee OA severity prediction in the past few years. Shamir \textit{et al.} \cite{Shamir} utilized a weighted nearest neighbors algorithm that incorporated the hand-crafted features like Gabor filters, Chebyshev statistics, multi-scale histograms, etc. Antony \textit{et al.} \cite{Antony16} proposed to utilize the transfer learning of the existing pre-trained deep CNNs. Later, Antony \textit{et al.} \cite{Antony17} customized a deep CNN from scratch and optimized the network using a weighted combination of the traditional cross-entropy and the mean squared error, which served as dual-objective learning. Tuilpin \textit{et al.} \cite{Tuilpin} developed a method inspired from the deep Siamese network \cite{Siamese}, for learning the similarity metric between the pair of radiographs. Gorriz \textit{et al.} \cite{gorriz19a} developed an end-to-end attention-based network, bypassing the need to localize knee joint, to quantify the knee OA severity automatically. Chen \textit{et al.} \cite{chen2019fully} proposed to utilize pre-trained VGG-19 \cite{simonyan2015deep} along with an adjustable ordinal loss for the proportionate penalty to the misclassification. Yong \textit{et al.} \cite{Yong} utilized the pretrained DenseNet-161 \cite{DenseNet}, along with an ordinal regression module (ORM), in order to treat the ordinality of the KL grading. They further optimized the network using the cumulative link (CL) loss function.  
     \begin{figure*}[t]
    \centering
    \resizebox{\textwidth}{!}{
    \includegraphics[scale = 0.25]{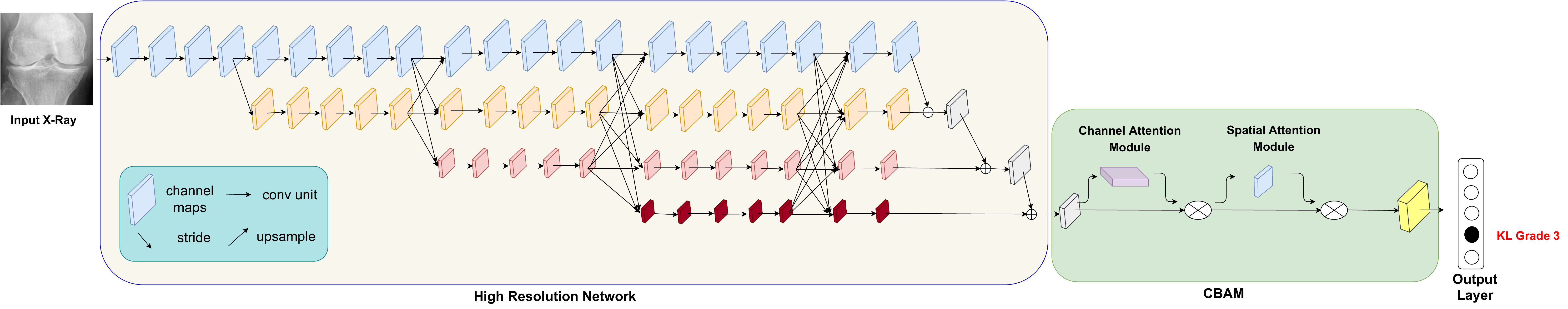}}
    \caption{An overview of the architecture of the proposed OsteoHRNet for the knee OA severity prediction. Blocks with different colors denote convolution features at different spatial scales. The proposed model takes knee X-Ray image as input and estimates the OA severity in terms of KL grade.}
    \label{fig:Architecture}
\end{figure*}
\subsection{Motivation}
Deep CNNs are renowned for learning the highly correlated features in an image. In addition, it is a widely known fact that the first few layers of a deep CNN contribute to the learning of low-level features in an image. Whereas the last few layers contribute to the learning of the high-level features, enabling the final classification by adaptively learning spatial hierarchies of features \cite{Yamashita}. While the low-level features are the minute details of an image, including points, lines, edges, etc., the high-level features comprise several low-level features, which make up the more prominent and robust structures for classification. 
    
However, in general, the knee X-Rays do not
comprise many edgy or low-level structures. Due to a lack of such vital information, it may be difficult for a deep CNN to learn an efficient classification particularly, in the case of knee OA, where one KL grade is not very distinctive from the other unless carefully inspected (see Fig. \ref{fig:Sample}). \ul{A few of the most recent state-of-the-art methods} \cite{chen2019fully}, \cite{Yong} \ul{have directly utilized the existing popular image classification models in a plug-and-play fashion without supervising the network engineering relevant to the given problem. It should be mentioned that while a majority of those methods were built for a generic image classification problem, a few of them were explicitly designed using architectural search, e.g., MobileNetV2} \cite{sandler2019mobilenetv2}. 
    
    Moreover, for the knee OA severity classification, the presented best-performing deep CNNs were enormous in size, exceeding 500 MB \cite{chen2019fully}, to be precise. As a result, such models may require substantially high computational resources, making it challenging to deploy in real-time environments. Therefore, it may be said that the direct usage of popular classification models may not be appropriate. Although some recent methods \cite{Tuilpin}, \cite{Zhang}, \cite{gorriz19a} \cite{Yong}, have started to design the models specific to knee OA given the amount of information present in the knee X-rays. However, they still lack in terms of accuracy and computational overhead. For \textit{e.g.,} Zhang \textit{et al.} \cite{Zhang} utilized the Convolutional Block Attention Module, namely CBAM \cite{CBAM}, after every residual layer in their proposed architecture, which may not be computationally pleasant. The attention module has performed undoubtedly well in many high-level vision tasks. However, one must not overlook its computational overhead considering the presence of fully connected layers. 

\begin{figure}[t]
  \centering
      \includegraphics[scale = 0.4]{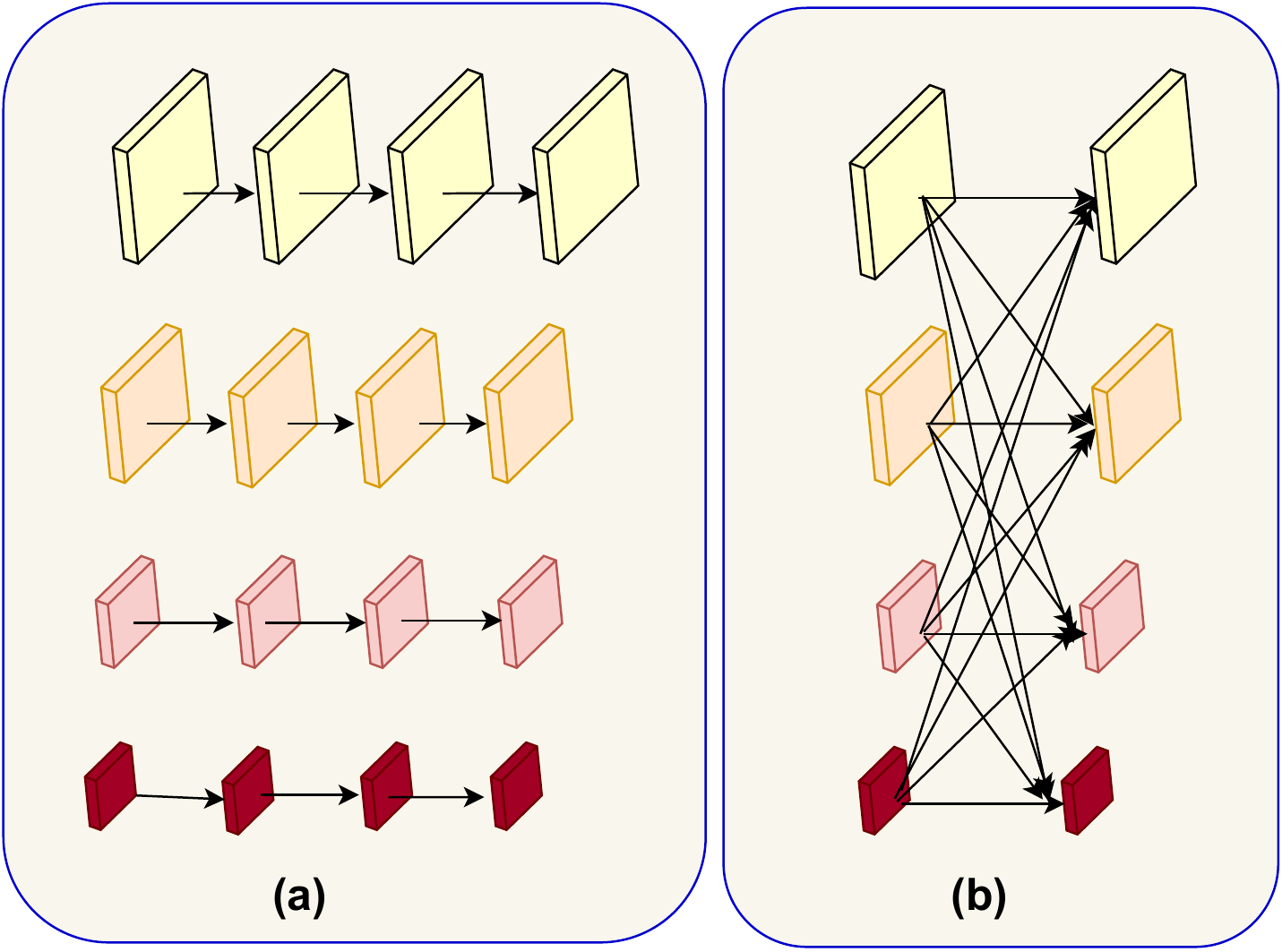}
    \caption{\textbf{Connections in HRNet: }(a) Multi Resolution convolution in parallel, (b) Fusion of Multi-Resolution convolution. Different colors denote feature resolution at various scales.}
    \label{fig:SubNetwork}
\end{figure}
    The applicability of deep CNNs in medical imaging heavily depends on the amount of data available for efficient learning. As an alternative, many deep learning-based methods have utilized the data augmentation techniques to further boost the performance, which has not been much considered in the existing works.
    
    \subsection{Our Contributions}    
Based on the aforementioned drawbacks of the existing best-published works, our contributions are five-fold, as follows:
\begin{enumerate}
    \item We propose an efficient deep CNN for the knee OA severity prediction in terms of KL grades using X-ray images. Unlike existing methods, our proposed scheme is not a blind plug-and-play of popular deep models. The proposed scheme has been built upon a high-resolution network; namely, HRNet \cite{HRNet}, that takes the spatial scale of the X-Ray image into account for efficient classification.
    
    \item We also propose to utilize the attention mechanism only once in the entire network to reduce the computational overhead and adaptive filtering of the counterproductive features just before classification.
    
    \item Also, instead of relying on traditional entropy-based minimization, we have adopted the ordinal loss \cite{chen2019fully} to optimize the proposed scheme.
    
    \item To further boost the performance of the proposed scheme, we have incorporated the data augmentation techniques, which have not been much considered in any recent work so far.

    \item  Lastly, we present an extensive set of experiments and Grad-CAM \cite{GRADCam} visualization to justify the importance of each module of the proposed framework.
\end{enumerate}

The rest of the paper is organized as follows: Section \ref{Proposed} presents the proposed method and the adopted cost function. Section \ref{Experiments} briefly describes the incorporated dataset, training details,  competing methods, and evaluation metrics. Section \ref{Results} presents the quantitative and qualitative comparison against the best-published works. Section \ref{Discussion} presents a brief discussion on the learning of proposed scheme in terms of Grad-CAM visualization of obtained results. Section \ref{Ablation} demonstrates the ablation study against various components, and finally, the paper is concluded in Section \ref{Conclusions}.  

\section{Proposed Method}
\label{Proposed}
This section presents the details of the proposed model, followed by a brief description of the incorporated cost function. The proposed framework is built upon the HRNet and Convolution Block Attention Module (CBAM) in a serially cascaded manner. A descriptive representation of the proposed model is shown in Fig. \ref{fig:Architecture}.

\subsection{High Resolution Network}
High-Resolution Network (HRNet) \cite{HRNet} is a novel and revolutionary multi-resolution deep CNN, which tends to maintain high-resolution feature representations throughout the network. It starts as a stream of 2D convolutions and subsequently adds up the high-to-low resolution streams to form the following stages. It then merges the multi-resolution streams in parallel for information exchange \cite{HRNet} as shown in Fig. \ref{fig:Architecture} (\textit{marked as High-Resolution Network}). HRNet tends to generate reliable multi-resolution representations with strong spatial sensitivity. It has been achieved by utilizing parallel connections instead of serial ({see Fig. \ref{fig:SubNetwork}(a)}) and recurrent fusion of the intermediate representations from multi-resolution streams ({see Fig. \ref{fig:SubNetwork}(b)}), as shown in Fig. \ref{fig:SubNetwork}. As a result, it enables the network to learn more highly correlated and semantically robust spatial features. This motivates us to incorporate HRNet for processing the knee X-Ray images, which lack such rich spatial features. 

To formally define, let $\mathcal{D}_{ij}$ denotes the sub-network in the $i^{th}$ stage of $j^{th}$ resolution index. The spatial resolution in this branch is $1/2^{j}-1$ of that of the high-resolution (HR) branch. For \textit{e.g.,} HRNet, which consists of four different resolution scales, can be illustrated as follows:
\begin{equation}
\begin{array}{llll}
\mathcal{D}_{11}  & \rightarrow ~~ \mathcal{D}_{21} &\rightarrow ~~ \mathcal{D}_{31} & \rightarrow ~~ \mathcal{D}_{41} \\
& \searrow ~~ \mathcal{D}_{22} &\rightarrow ~~ \mathcal{D}_{32} & \rightarrow ~~ \mathcal{D}_{42} \\
&   &\searrow  ~~ \mathcal{D}_{33} & \rightarrow ~~ \mathcal{D}_{43}  \\
&   & & \searrow ~~ \mathcal{D}_{44},
\end{array}
\end{equation}

\begin{figure}[t]
    \centering
    \includegraphics[width=0.45\textwidth]{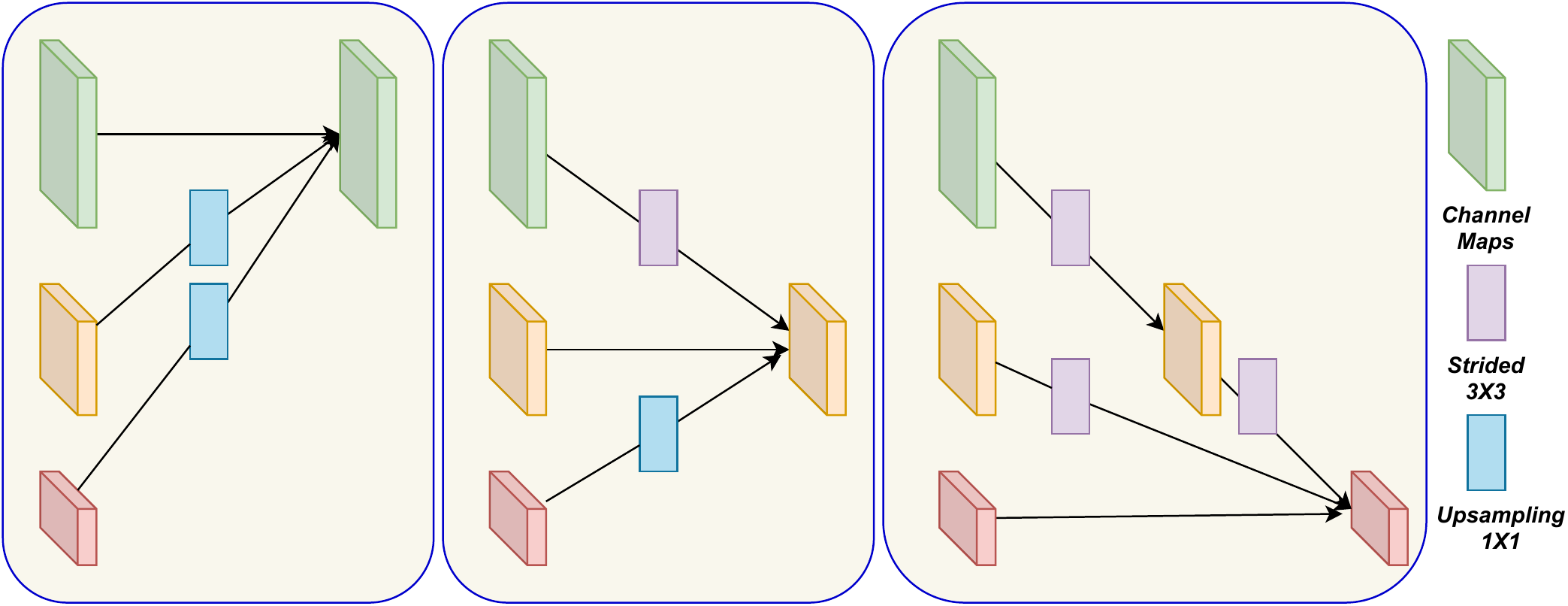}
    \caption{Graphical demonstration of how HRNet fuses information from different resolutions.}
    \label{fig:fusion_case}
\end{figure}
Later, the obtained multi-resolution feature maps are fused to exchange the learned variscaled information, as shown in Fig. \ref{fig:fusion_case}. For this, HRNet utilizes bilinear upsampling followed by the $1\times 1$ convolution to adjust the number of channels when transforming the lower resolution feature map to a higher resolution scale, or a strided $3\times 3$ convolution otherwise.

\subsection{Convolutional Block Attention Module}
Convolutional Block Attention Module (CBAM) consists of two sequential sub-modules : (a) channel attention module, and (b) spatial attention module \cite{CBAM}.
Given an input feature map, $\mathbf{P} \in \mathbb{R}^{C \times H \times W}$, CBAM sequentially infers a one-dimensional channel attention map ${Map}_{c} \in  \mathbb{R}^{C\times 1\times 1}$ and a two-dimensional spatial attention map ${Map}_{s}   \in \mathbb{R}^{1\times H \times W}$. Thus we obtain a final refined attention map, here denoted as \textbf{T}, and the comprehensive attention mechanism can be summarized as:
\begin{equation}
\begin{split}
    \mathbf{P}^{c} &= Map_c (\mathbf{P}) \otimes \mathbf{P},\\
    \mathbf{T} &= Map_s (\mathbf{P}^c) \otimes \mathbf{P}^c,
\end{split}
\end{equation}
where $\otimes$ signifies element-wise multiplication. $Map_c$ is first generated by making use of the cross-channel relationship of the features, as,
\begin{equation}
    Map_c (\mathbf{P}) = g(MLP(\mathcal{A}(\mathbf{P}))) +  MLP(\mathcal{M}(\mathbf{P})),
\end{equation}
where $g$, $MLP$, $\mathcal{A}$, and $\mathcal{M}$ denote sigmoid function, multi-layer perceptron, average pool and max pool, respectively.

Whereas, the $Map_s$ is generated efficiently by performing $\mathcal{M}$ and $\mathcal{A}$ along the channel axis. Next, the pooled descriptors are concatenated together to generate a reliable and efficient feature descriptor by utilizing the inter-spatial correlation of the features. It can be written as,
\begin{equation}
    Map_s (\mathbf{P}) = g(k^{7\times 7}([\mathcal{A}(\mathbf{P}); \mathcal{M}(\mathbf{P})])),
\end{equation}
where $k^{7\times 7}$ denotes the convolution operation with kernel of size $7 \times 7$.
\subsection{Network Architecture}
We propose a deep CNN, called OsteoHRNet, that utilizes the HRNet as the backbone and is further empowered with an attention mechanism for the knee KL grade classification. CBAM is integrated at the end of the HRNet, followed by a fully connected (FC) output layer, as depicted in Fig. \ref{fig:Architecture}. It may be said that the integration of the CBAM module after HRNet has been beneficial in learning adaptive enriched features for an efficient KL grade classification. It can also be observed that the proposed one-time integration of CBAM is computationally pleasant, compared to the multiple additions in the existing work \cite{Zhang}. The resultant output from the CBAM is then fed into the final fully connected layer, which outputs the probabilities of the KL grade for the given input X-Ray image. HRNet has been considered for reliable feature extraction, whereas the capabilities of CBAM are leveraged to help the model better focus on relevant features.     

\subsection{Cost Functions}
A majority of the existing works on knee OA severity classification have considered the nominal nature of KL grades for classification. However, inspired by the idea of Chen \textit{et al.} \cite{chen2019fully}, we approach this task as an ordinal regression problem and therefore utilize the ordinal loss function instead of the traditional cross-entropy. The ordinal loss function used in this paper is a weighted ratio of the traditional cross-entropy. Given the ordinality in the KL grading, it must be acknowledged that extra information is provided by progressive grading. This approach penalizes the distant grade misclassification more than the nearby grade according to the penalty weights. For \textit{e.g.,} a grade 1 classified as grade 3 is penalized more severely than it is classified as grade 2 and even more for being classified as grade 4. An ordinal matrix \textbf{$C_{n \times n}$} is considered as the penalty weights between the outcome and the true grade, i.e.,  \textbf{$c_{uv}$} denotes the penalty weight for predicting a grade \textit{v} as \textit{u} with $n = 5$. In this study, with five KL grades to classify and $c_{uu} = 1$, the adopted ordinal loss can be written as
\begin{equation}
    \centering
    \mathcal{L}_{o} = \sum_{u=0} ^{n-1} c_{uv} * q_{u},
\end{equation}
where $u$, $v$ are the predicted and true KL grades of the input image, respectively, $p_u$ is the output probability by the final output layer of the architecture with $q_{u} = p_{u}$ if \textit{u $\neq$ v} and  $q_{u} = 1-p_{u}$ , otherwise. We have utilized the following penalty matrix for our experimentation.
\begin{equation*}
\begin{bmatrix}
1 & 3 & 6 & 7 & 9\\
4 & 1 & 4 & 5 & 7\\
6 & 4 & 1 & 3 & 5\\
9 & 7 & 4 & 1 & 4\\
11 & 9 & 7 & 5 & 1\\
\end{bmatrix}    
\end{equation*}

\section{Experimental Details} \label{Experiments}
\subsection{Dataset}
We have utilized the X-ray radiographs acquired from the OAI repository which has been made available by Chen \textit{et al.} \cite{dataset}. The images obtained are of 4796 participants, including men and women. Given that we focus primarily on the KL grades, radiographs with annotated KL grades from the baseline cohort are acquired to assess our method. The dataset of a total of 8260 radiographs, including the left and right knee, was split into train, test, and validation sets in the ratio of 7:2:1 with balanced distribution across all KL grades \cite{dataset}. Table \ref{table:Distribution} shows the train, test, and validation distribution of the dataset.  
\begin{table}[t] 
  \centering
  \caption{Distribution of the dataset}
  \label{table:Distribution}
   \resizebox{0.47\textwidth}{!}{\begin{tabular}{||c|| c c c c c ||c||} 
 \hline
 Dataset & Grade0 & Grade1 & Grade2 & Grade3 & Grade4 & Total \\ [0.5ex] 
 \hline\hline
Training & 2286 & 1046 & 1516 & 757 & 173 & 5778\\ [0.5ex]

 \hline
Testing & 639 & 296 & 447 & 223 & 51 & 1656 \\ [0.5ex]
 \hline
Validation & 328 & 153 & 212 & 106 & 27 & 826\\ [0.5ex]
 \hline
Total & 3253 & 1495 & 2175 & 1086 & 251 & 8260\\ [0.5ex]
 \hline
 \end{tabular}}
\end{table}

\begin{figure*}[!hbt]
    \centering
    \resizebox{\textwidth}{!}{
    \setlength{\tabcolsep}{1pt}
    \begin{tabular}{ccc}
         \includegraphics{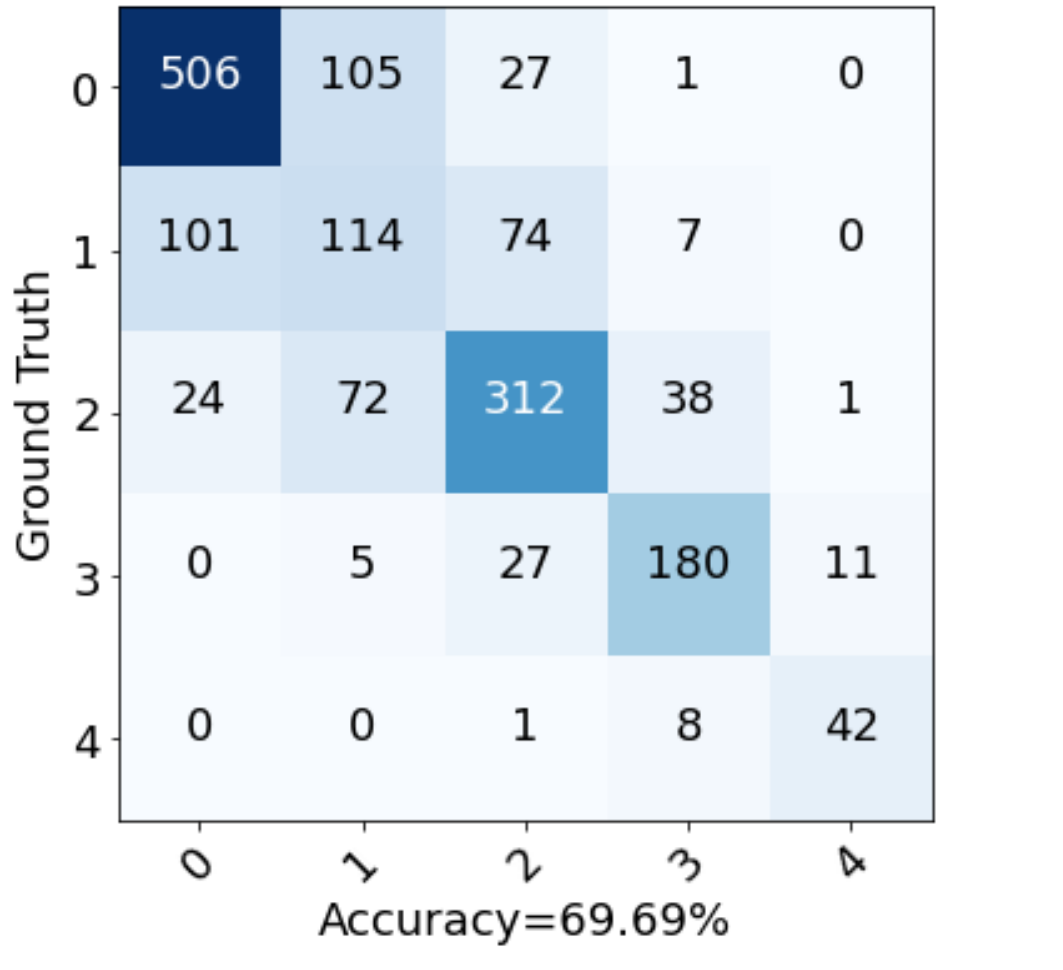}&
         \includegraphics{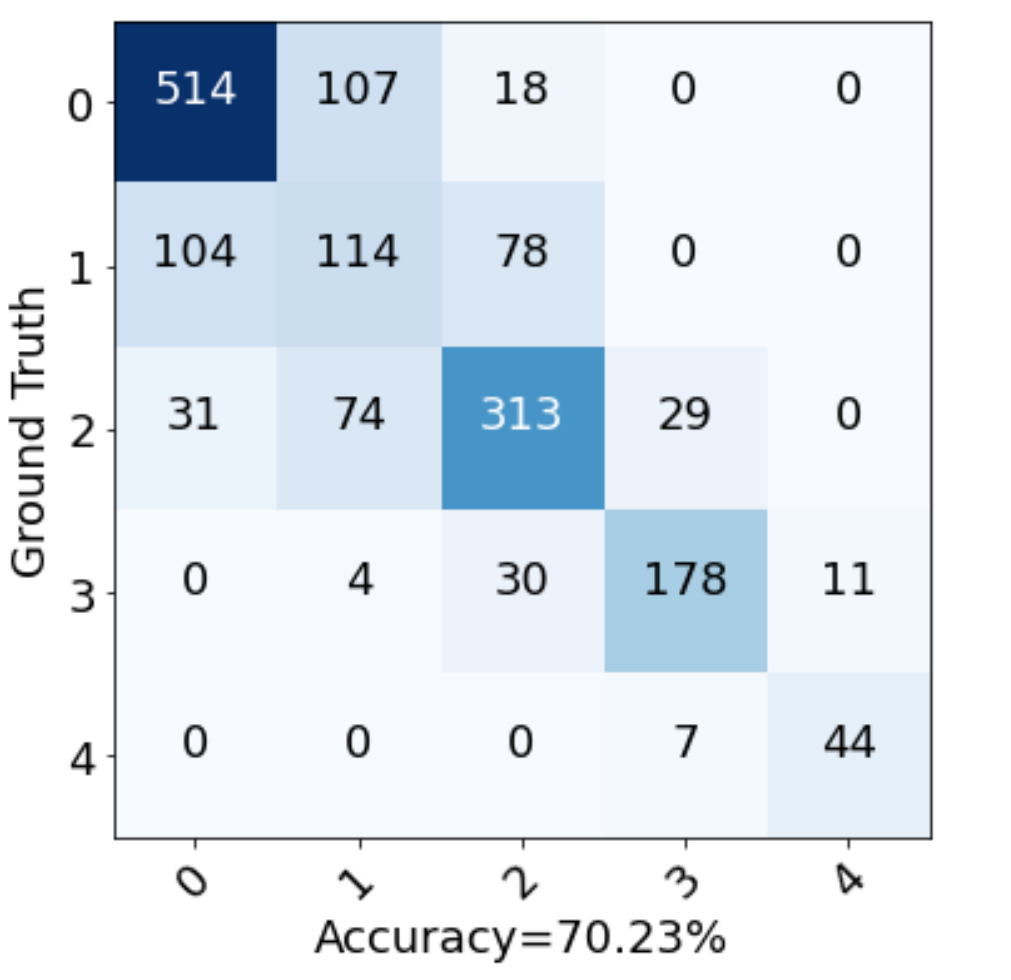}&
         \includegraphics{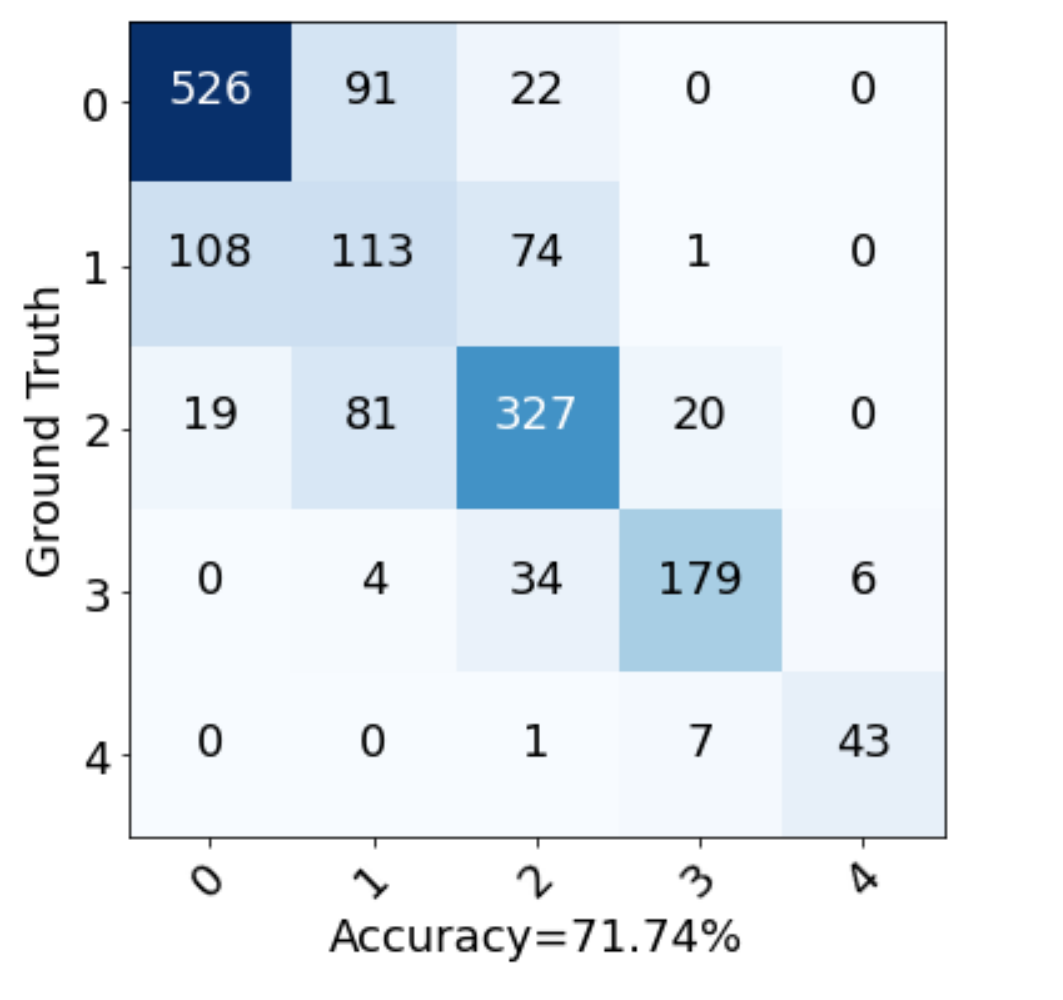}\\
         Chen \textit{et al.} \cite{chen2019fully} & Yong \textit{et al.} \cite{Yong} & OsteoHRNet\\
    \end{tabular}
    }
    \caption{Confusion matrices for KL grade prediction using different competing approaches \cite{chen2019fully}, \cite{Yong} and OsteoHRNet.}
    \label{fig:ConfusionMatrix}
\end{figure*}

\begin{figure*} [h] 
  \centering
  \setlength{\tabcolsep}{1pt}
  \resizebox{1\textwidth}{!}{
  \begin{tabular}{!{\color{blue}\vrule}cccccccc!{\color{blue}\vrule}} 
        \arrayrulecolor{blue}\hline
    \multicolumn{2}{|c}{\includegraphics[width=3cm, height=3cm]{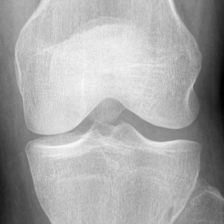}} &
    \multicolumn{2}{c}{\includegraphics[width=3cm, height=3cm]{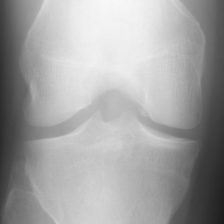}} &
    \multicolumn{2}{c}{\includegraphics[width=3cm, height=3cm]{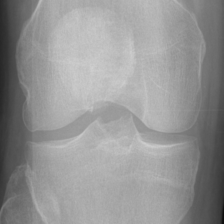}} &
    \multicolumn{2}{c|}{\includegraphics[width=3cm, height=3cm]{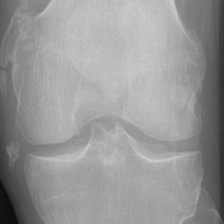}}\\
    
    \includegraphics[width=3cm, height=3cm,trim=1.2cm 0.3cm 1.2cm 0.3cm,clip]{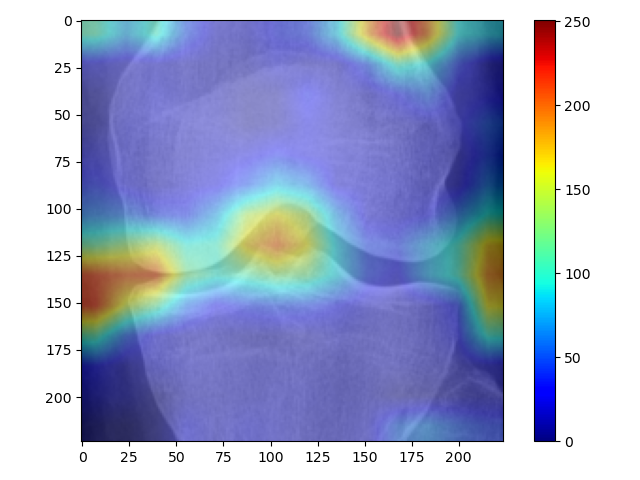} &
    \includegraphics[width=3cm, height=3cm,trim=1.2cm 0.3cm 1.2cm 0.3cm,clip]{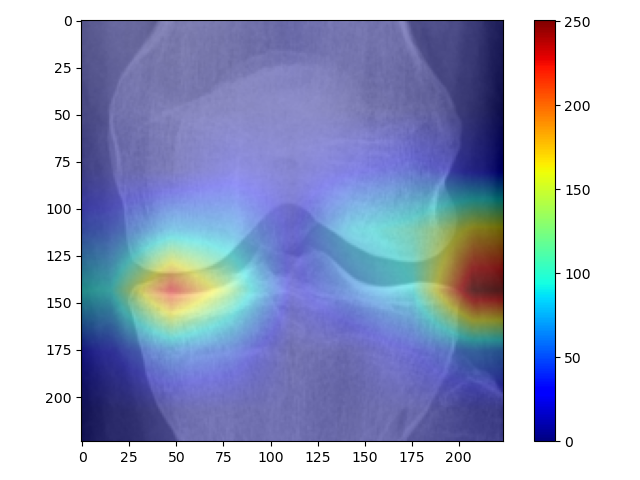} &
    \includegraphics[width=3cm, height=3cm,trim=1.2cm 0.3cm 1.2cm 0.3cm,clip]{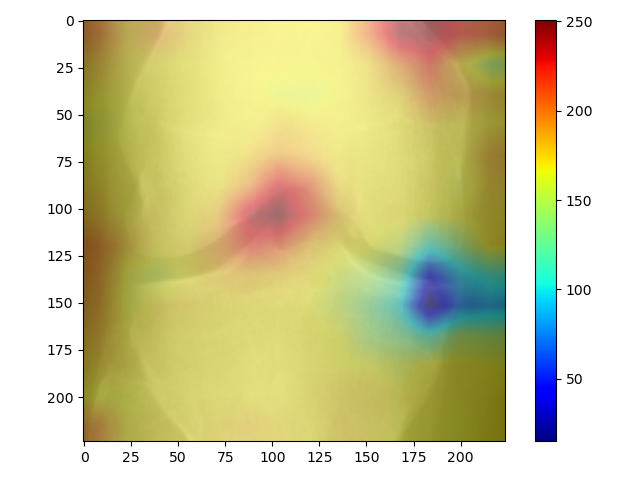} &
    \includegraphics[width=3cm, height=3cm,trim=1.2cm 0.3cm 1.2cm 0.3cm,clip]{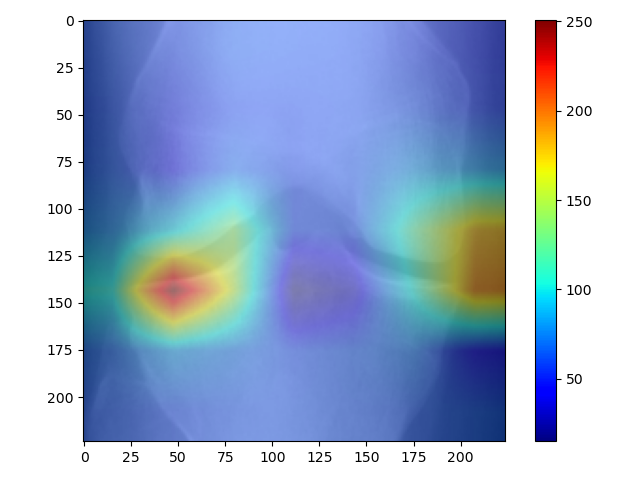} &
    \includegraphics[width=3cm, height=3cm,trim=1.2cm 0.3cm 1.2cm 0.3cm,clip]{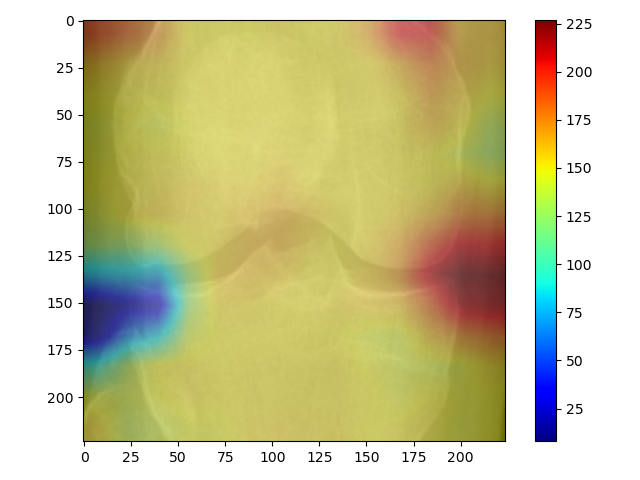} &
    \includegraphics[width=3cm, height=3cm,trim=1.2cm 0.3cm 1.2cm 0.3cm,clip]{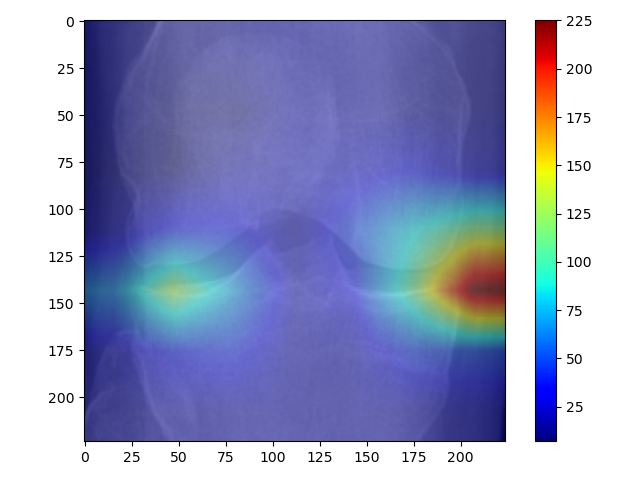} &
     \includegraphics[width=3cm, height=3cm,trim=1.2cm 0.3cm 1.2cm 0.3cm,clip]{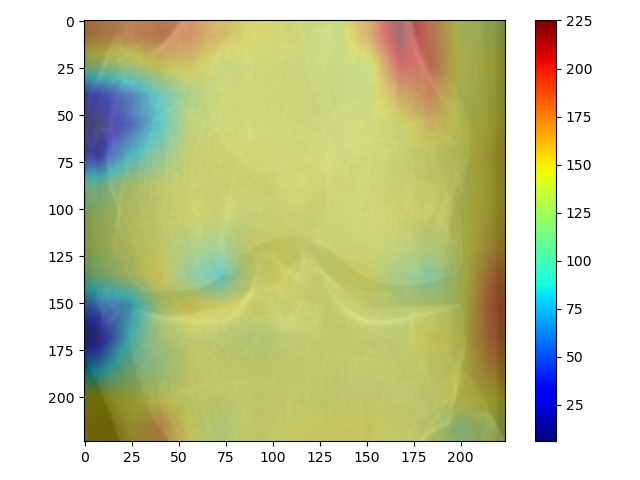} &
    \includegraphics[width=3cm, height=3cm,trim=1.2cm 0.3cm 1.2cm 0.3cm,clip]{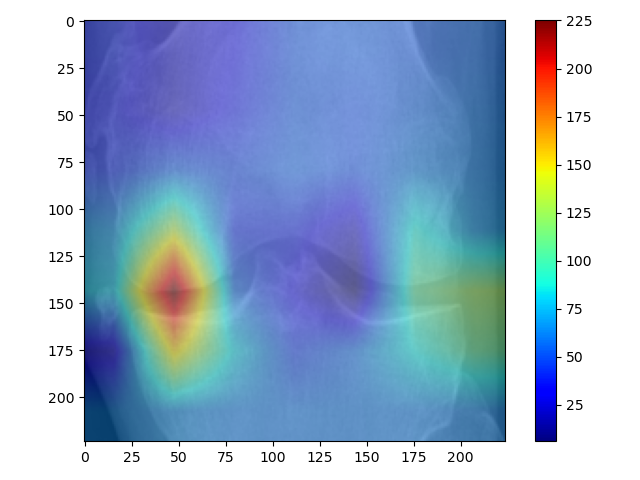}\\
   VGG-19 \cite{chen2019fully}& OsteoHRNet & VGG-19 \cite{chen2019fully}& OsteoHRNet & VGG-19 \cite{chen2019fully}& OsteoHRNet & VGG-19 \cite{chen2019fully}& OsteoHRNet\\
        \arrayrulecolor{blue}\hline
      \end{tabular}}
  \caption{Grad-CAM visualizations generated against KL grade 0 test images using Chen \textit{et al.} \cite{chen2019fully} and OsteoHRNet.} \label{fig:GradCam0}
\end{figure*}

\begin{figure*} [h] 
  \centering
  \setlength{\tabcolsep}{1pt}
  \resizebox{1\textwidth}{!}{
  \begin{tabular}{!{\color{blue}\vrule}cccccccc!{\color{blue}\vrule}} 
  
  \arrayrulecolor{blue}\hline
    \multicolumn{2}{|c}{\includegraphics[width=3cm, height=3cm]{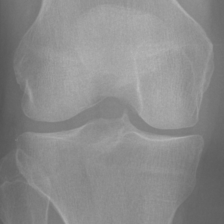}} &
    \multicolumn{2}{c}{\includegraphics[width=3cm, height=3cm]{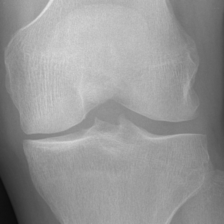}} &
    \multicolumn{2}{c}{\includegraphics[width=3cm, height=3cm]{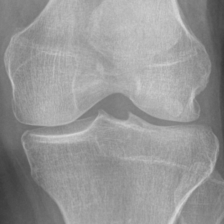}} &
    \multicolumn{2}{c|}{\includegraphics[width=3cm, height=3cm]{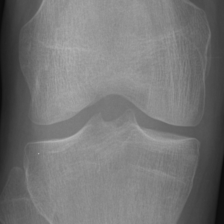}}\\

    \includegraphics[width=3cm, height=3cm,trim=1.2cm 0.3cm 1.2cm 0.3cm,clip]{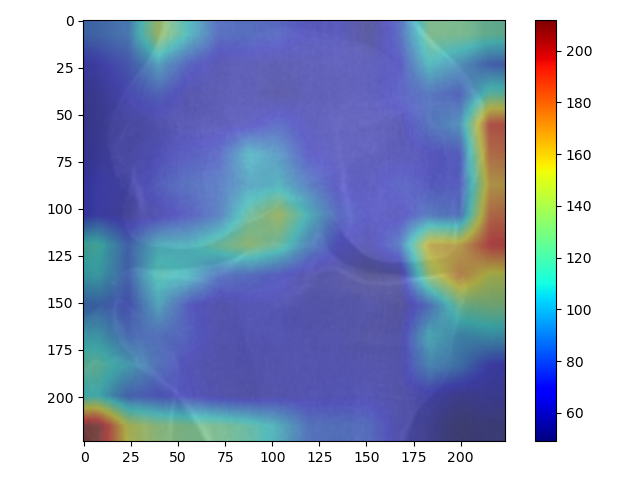} &
    \includegraphics[width=3cm, height=3cm,trim=1.2cm 0.3cm 1.2cm 0.3cm,clip]{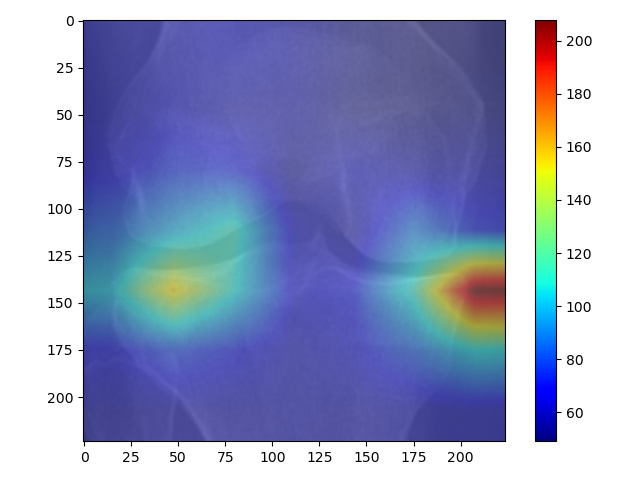} &
    \includegraphics[width=3cm, height=3cm,trim=1.2cm 0.3cm 1.2cm 0.3cm,clip]{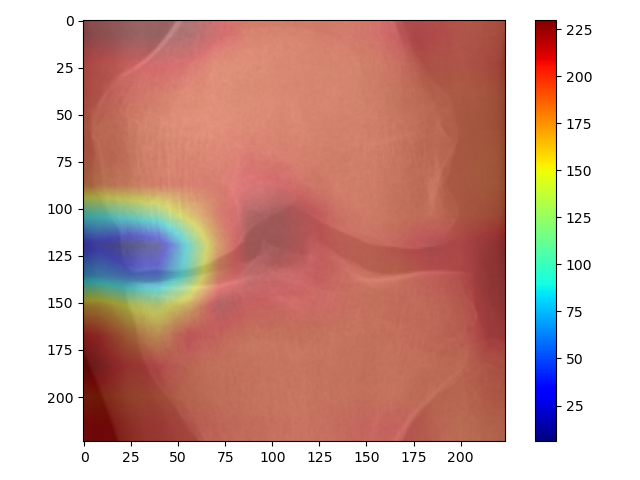} &
    \includegraphics[width=3cm, height=3cm,trim=1.2cm 0.3cm 1.2cm 0.3cm,clip]{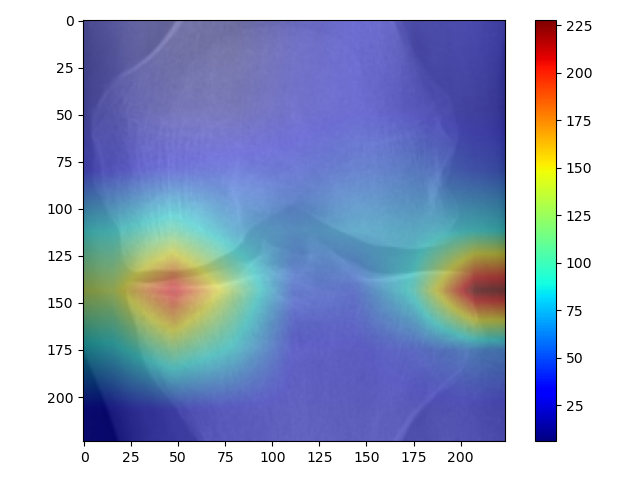} &
    \includegraphics[width=3cm, height=3cm,trim=1.2cm 0.3cm 1.2cm 0.3cm,clip]{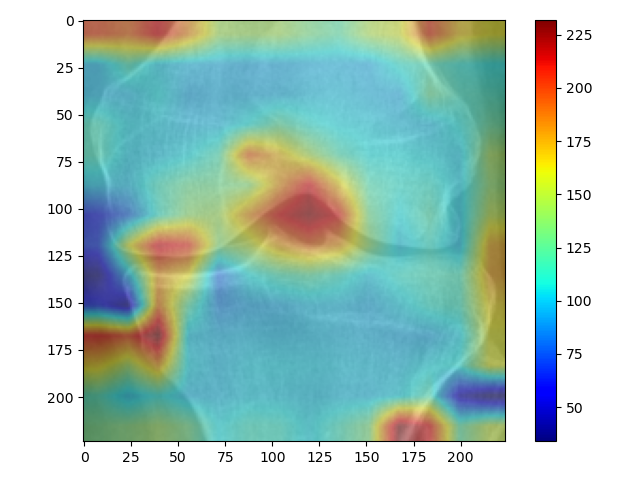} &
    \includegraphics[width=3cm, height=3cm,trim=1.2cm 0.3cm 1.2cm 0.3cm,clip]{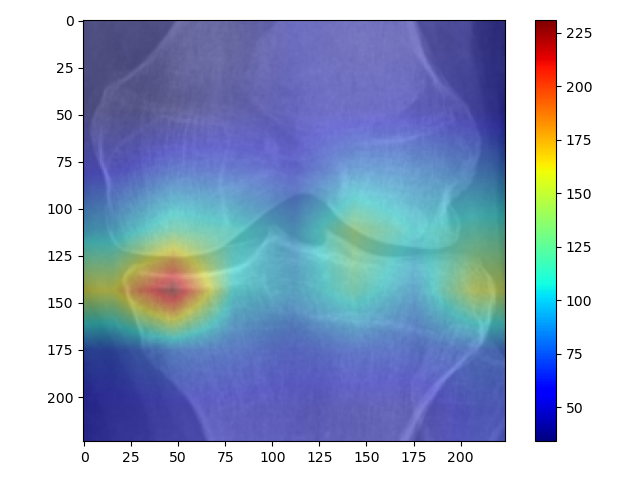} &
     \includegraphics[width=3cm, height=3cm,trim=1.2cm 0.3cm 1.2cm 0.3cm,clip]{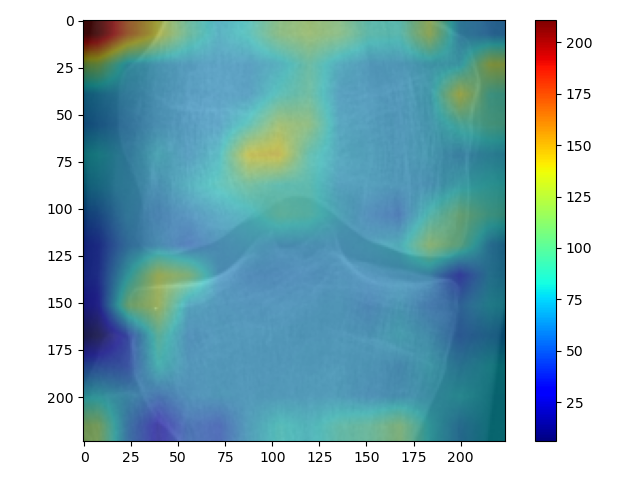} &
    \includegraphics[width=3cm, height=3cm,trim=1.2cm 0.3cm 1.2cm 0.3cm,clip]{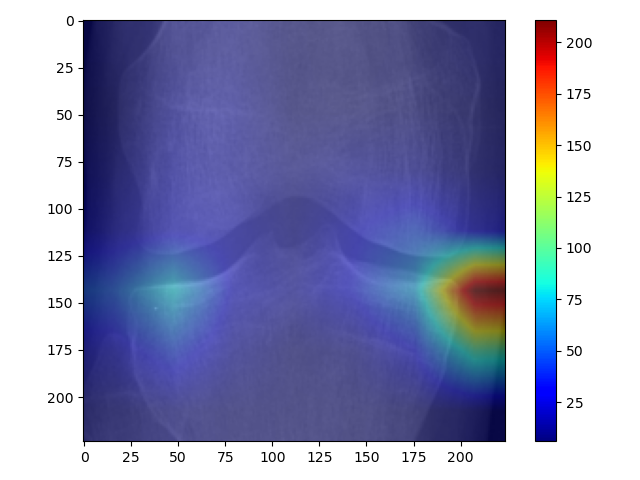}\\
     VGG-19 \cite{chen2019fully}& OsteoHRNet & VGG-19 \cite{chen2019fully}& OsteoHRNet & VGG-19 \cite{chen2019fully}& OsteoHRNet & VGG-19 \cite{chen2019fully}& OsteoHRNet\\

  \arrayrulecolor{blue}\hline
  
      \end{tabular}}
  \caption{Grad-CAM visualizations generated against KL grade 1 test images using Chen \textit{et al.} \cite{chen2019fully} and OsteoHRNet.} \label{fig:GradCam1}
\end{figure*}

\begin{figure*} [h] 
  \centering
  \setlength{\tabcolsep}{1pt}
  \resizebox{1\textwidth}{!}{
  \begin{tabular}{!{\color{blue}\vrule}cccccccc!{\color{blue}\vrule}} 
  \arrayrulecolor{blue}\hline
  \multicolumn{2}{|c}{\includegraphics[width=3cm, height=3cm]{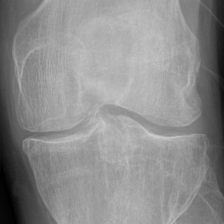}} &
    \multicolumn{2}{c}{\includegraphics[width=3cm, height=3cm]{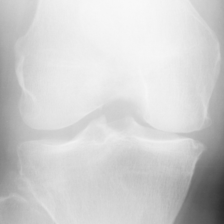}} &
    \multicolumn{2}{c}{\includegraphics[width=3cm, height=3cm]{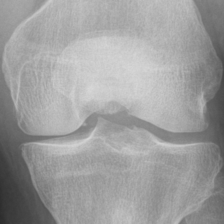}} &
    \multicolumn{2}{c|}{\includegraphics[width=3cm, height=3cm]{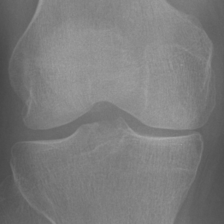}}\\

    \includegraphics[width=3cm, height=3cm,trim=1.2cm 0.3cm 1.2cm 0.3cm,clip]{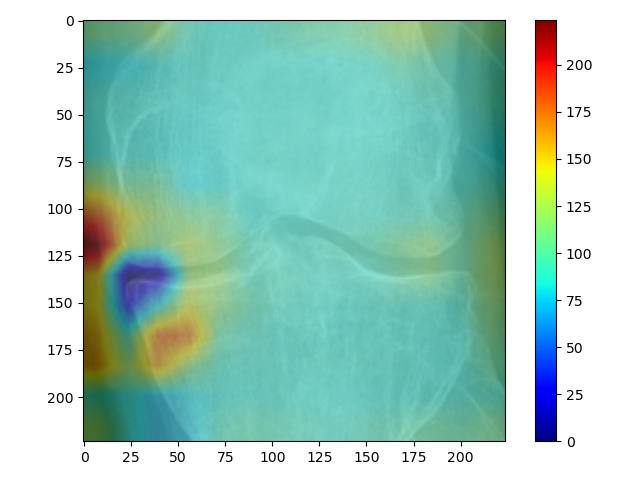} &
    \includegraphics[width=3cm, height=3cm,trim=1.2cm 0.3cm 1.2cm 0.3cm,clip]{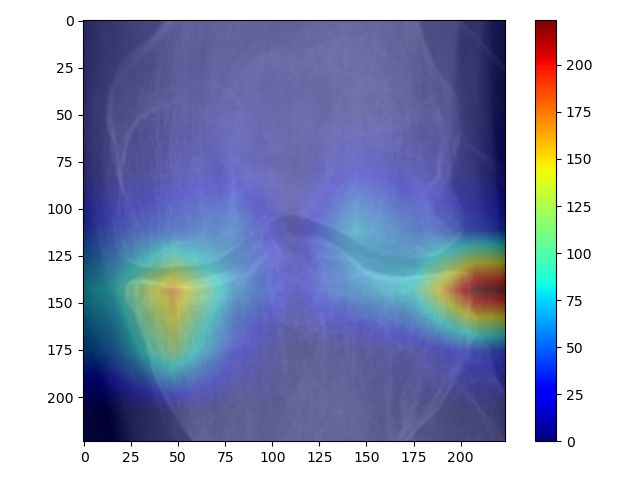} &
    \includegraphics[width=3cm, height=3cm,trim=1.2cm 0.3cm 1.2cm 0.3cm,clip]{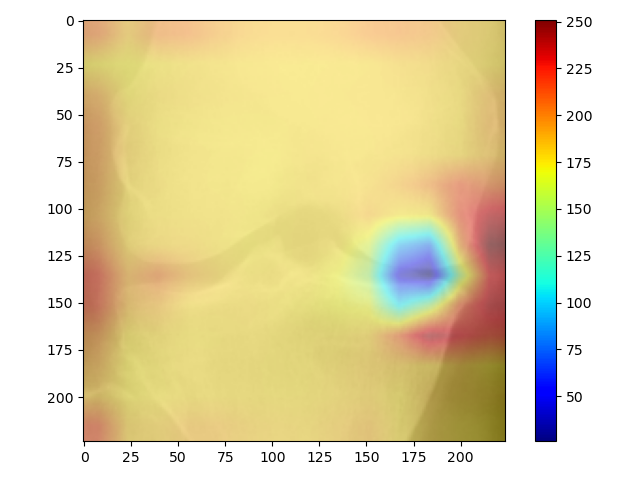} &
    \includegraphics[width=3cm, height=3cm,trim=1.2cm 0.3cm 1.2cm 0.3cm,clip]{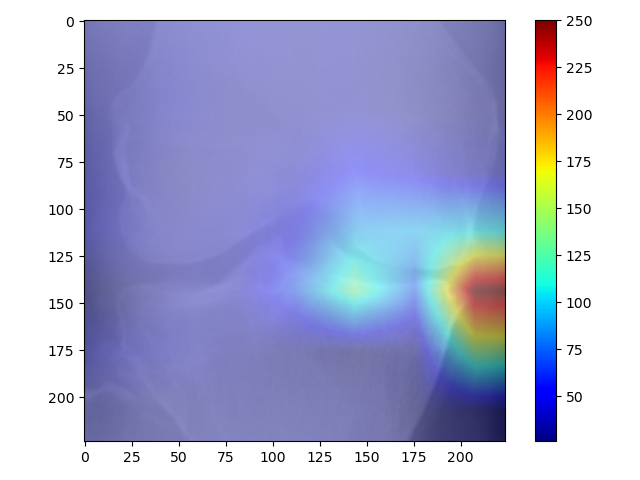} &
    \includegraphics[width=3cm, height=3cm,trim=1.2cm 0.3cm 1.2cm 0.3cm,clip]{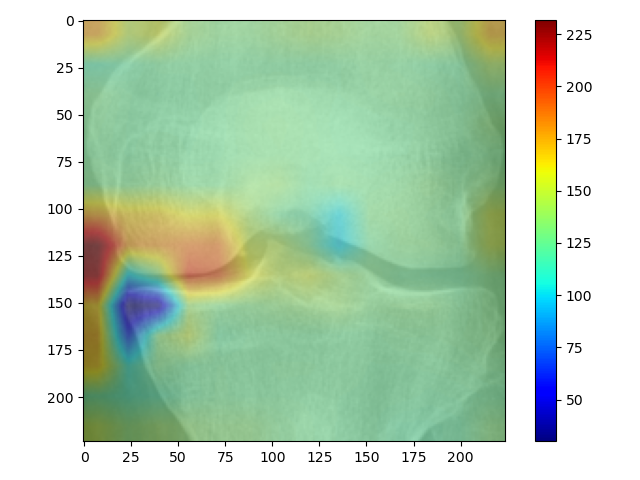} &
    \includegraphics[width=3cm, height=3cm,trim=1.2cm 0.3cm 1.2cm 0.3cm,clip]{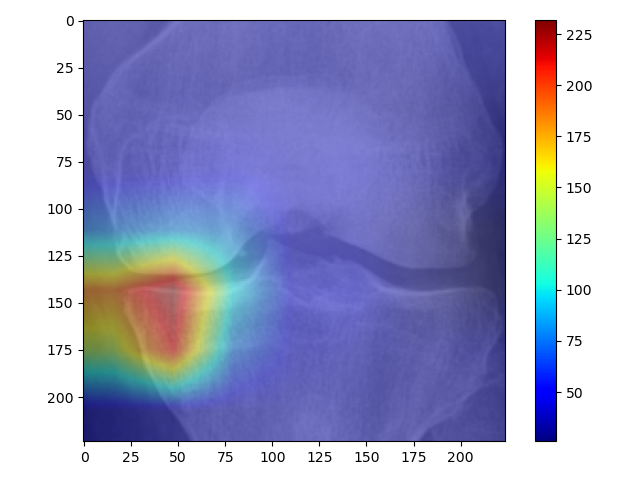} &
     \includegraphics[width=3cm, height=3cm,trim=1.2cm 0.3cm 1.2cm 0.3cm,clip]{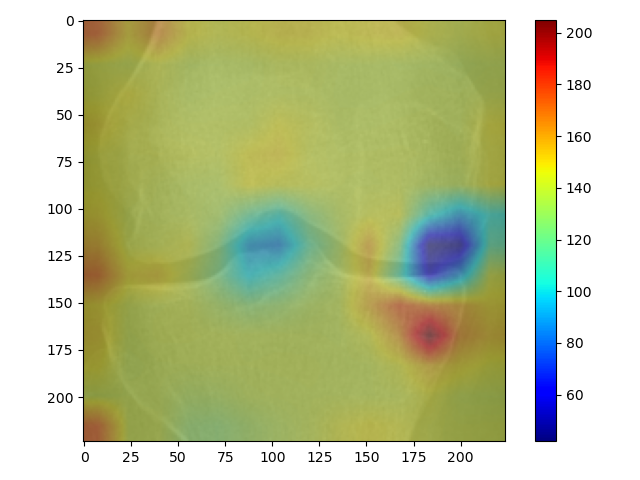} &
    \includegraphics[width=3cm, height=3cm,trim=1.2cm 0.3cm 1.2cm 0.3cm,clip]{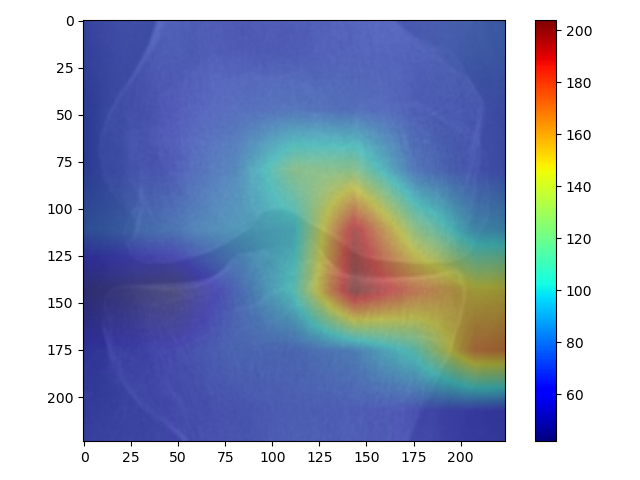}\\
     VGG-19 \cite{chen2019fully}& OsteoHRNet & VGG-19 \cite{chen2019fully}& OsteoHRNet & VGG-19 \cite{chen2019fully}& OsteoHRNet & VGG-19 \cite{chen2019fully}& OsteoHRNet\\
\arrayrulecolor{blue}\hline
      \end{tabular}}
  \caption{Grad-CAM visualizations generated against KL grade 2 test images using Chen \textit{et al.} \cite{chen2019fully} and OsteoHRNet.} \label{fig:GradCam2}
\end{figure*}

\begin{figure*} [h] 
  \centering
  \setlength{\tabcolsep}{1pt}
  \resizebox{1\textwidth}{!}{
  \begin{tabular}{!{\color{blue}\vrule}cccccccc!{\color{blue}\vrule}} 
    \arrayrulecolor{blue}\hline
    \multicolumn{2}{|c}{\includegraphics[width=3cm, height=3cm]{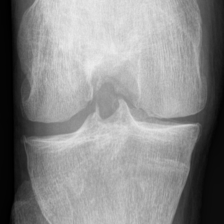}} &
    \multicolumn{2}{c}{\includegraphics[width=3cm, height=3cm]{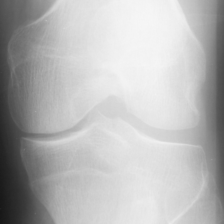}} &
    \multicolumn{2}{c}{\includegraphics[width=3cm, height=3cm]{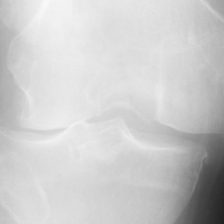}} &
    \multicolumn{2}{c|}{\includegraphics[width=3cm, height=3cm]{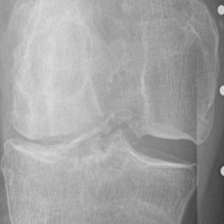}}\\

    \includegraphics[width=3cm, height=3cm,trim=1.2cm 0.3cm 1.2cm 0.3cm,clip]{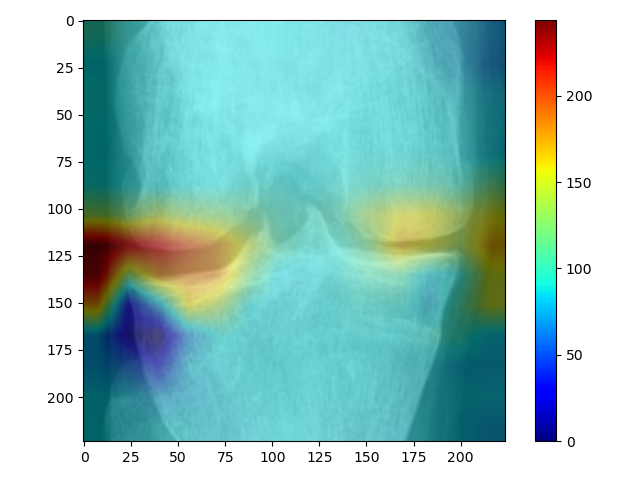} &
    \includegraphics[width=3cm, height=3cm,trim=1.2cm 0.3cm 1.2cm 0.3cm,clip]{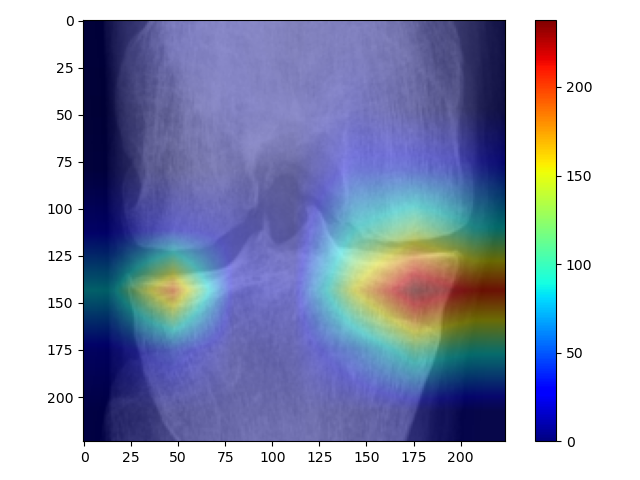} &
    \includegraphics[width=3cm, height=3cm,trim=1.2cm 0.3cm 1.2cm 0.3cm,clip]{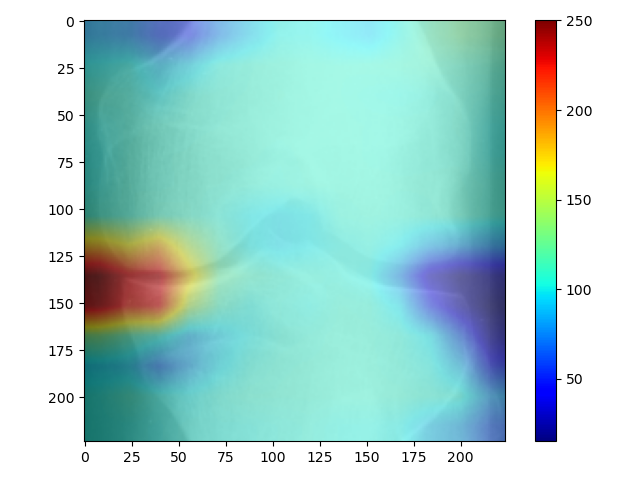} &
    \includegraphics[width=3cm, height=3cm,trim=1.2cm 0.3cm 1.2cm 0.3cm,clip]{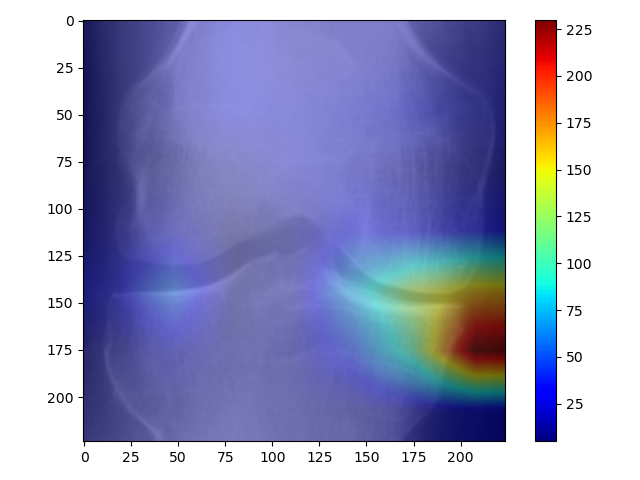} &
    \includegraphics[width=3cm, height=3cm,trim=1.2cm 0.3cm 1.2cm 0.3cm,clip]{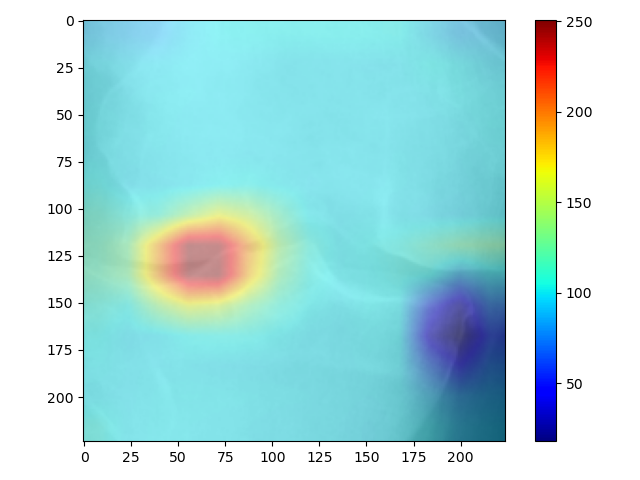} &
    \includegraphics[width=3cm, height=3cm,trim=1.2cm 0.3cm 1.2cm 0.3cm,clip]{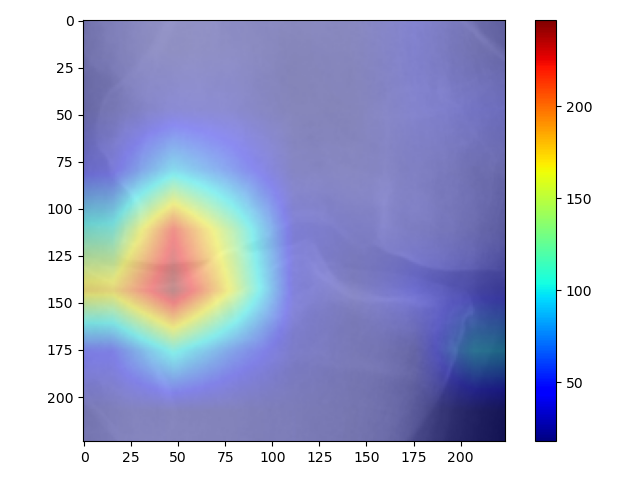} &
     \includegraphics[width=3cm, height=3cm,trim=1.2cm 0.3cm 1.2cm 0.3cm,clip]{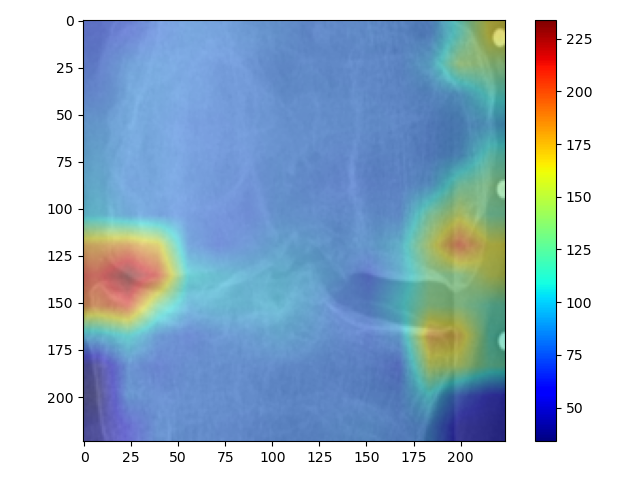} &
    \includegraphics[width=3cm, height=3cm,trim=1.2cm 0.3cm 1.2cm 0.3cm,clip]{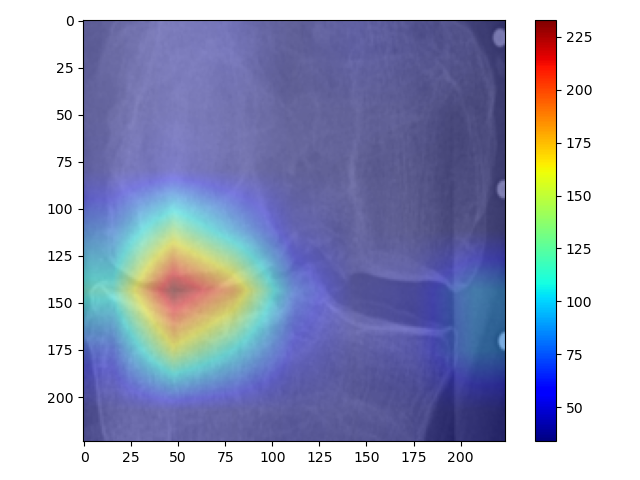}\\
     VGG-19 \cite{chen2019fully}& OsteoHRNet & VGG-19 \cite{chen2019fully}& OsteoHRNet & VGG-19 \cite{chen2019fully}& OsteoHRNet & VGG-19 \cite{chen2019fully}& OsteoHRNet\\

  \arrayrulecolor{blue}\hline
      \end{tabular}}
  \caption{Grad-CAM visualizations generated against KL grade 3 test images using Chen \textit{et al.} \cite{chen2019fully} and OsteoHRNet.} \label{fig:GradCam3}
\end{figure*}

\begin{figure*} [h] 
  \centering
  \setlength{\tabcolsep}{1pt}
  \resizebox{1\textwidth}{!}{
  \begin{tabular}{!{\color{blue}\vrule}cccccccc!{\color{blue}\vrule}} 
  \arrayrulecolor{blue}\hline
  \multicolumn{2}{|c}{\includegraphics[width=3cm, height=3cm]{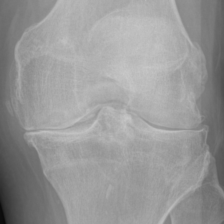}} &
    \multicolumn{2}{c}{\includegraphics[width=3cm, height=3cm]{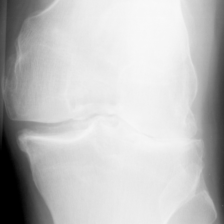}} &
    \multicolumn{2}{c}{\includegraphics[width=3cm, height=3cm]{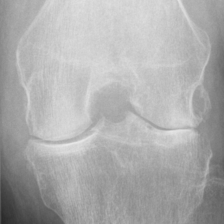}} &
    \multicolumn{2}{c|}{\includegraphics[width=3cm, height=3cm]{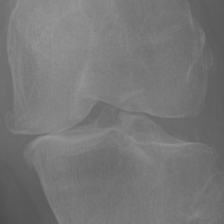}}\\

    \includegraphics[width=3cm, height=3cm,trim=1.2cm 0.3cm 1.2cm 0.3cm,clip]{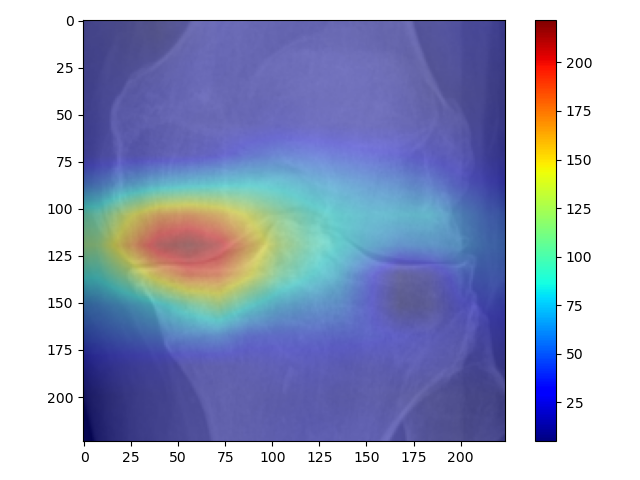} &
    \includegraphics[width=3cm, height=3cm,trim=1.2cm 0.3cm 1.2cm 0.3cm,clip]{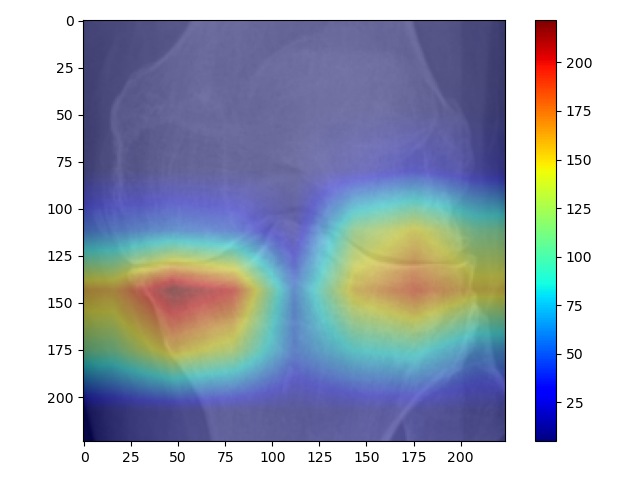} &
    \includegraphics[width=3cm, height=3cm,trim=1.2cm 0.3cm 1.2cm 0.3cm,clip]{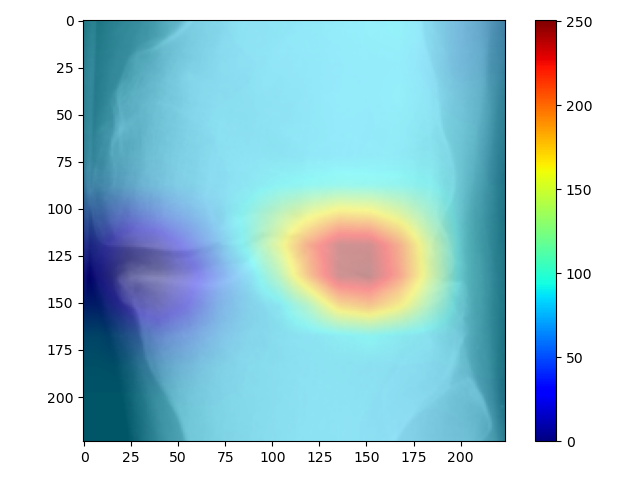} &
    \includegraphics[width=3cm, height=3cm,trim=1.2cm 0.3cm 1.2cm 0.3cm,clip]{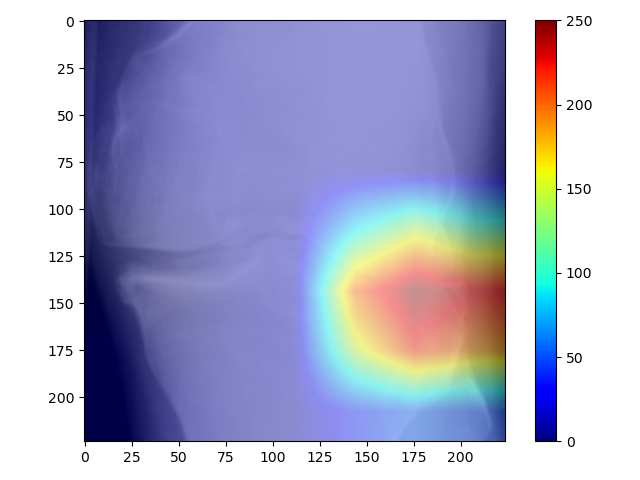} &
    \includegraphics[width=3cm, height=3cm,trim=1.2cm 0.3cm 1.2cm 0.3cm,clip]{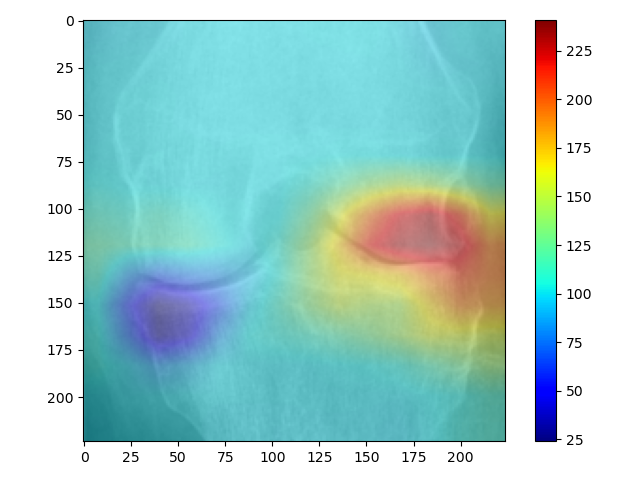} &
    \includegraphics[width=3cm, height=3cm,trim=1.2cm 0.3cm 1.2cm 0.3cm,clip]{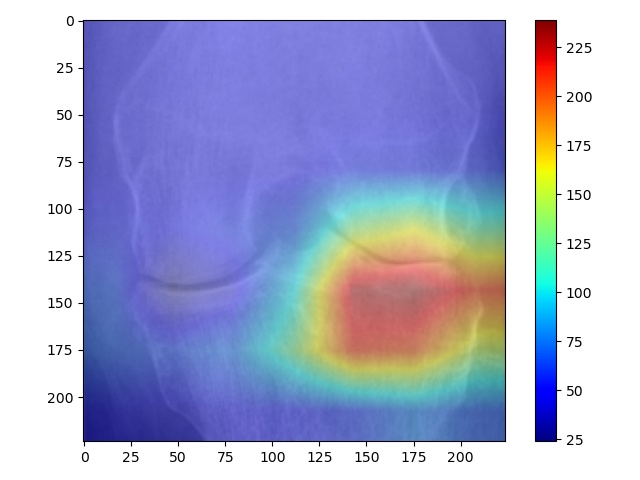} &
     \includegraphics[width=3cm, height=3cm,trim=1.2cm 0.3cm 1.2cm 0.3cm,clip]{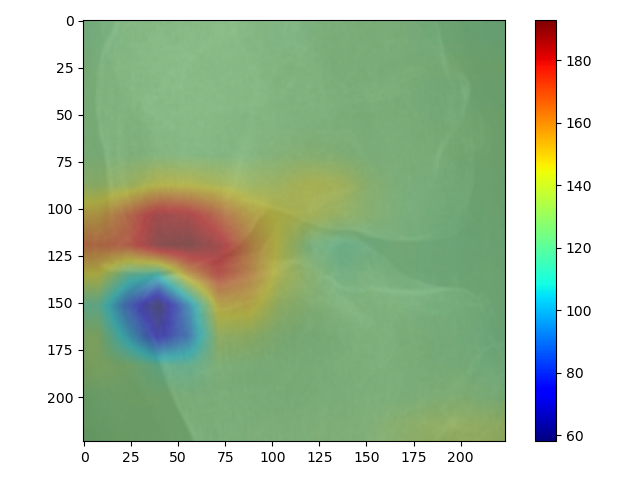} &
    \includegraphics[width=3cm, height=3cm,trim=1.2cm 0.3cm 1.2cm 0.3cm,clip]{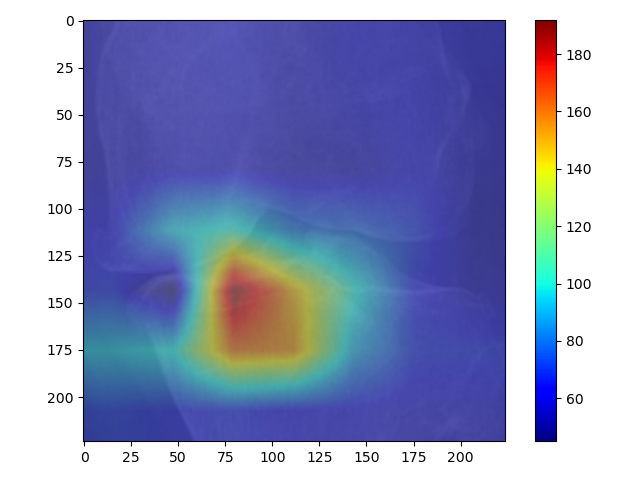}\\

     VGG-19 \cite{chen2019fully}& OsteoHRNet & VGG-19 \cite{chen2019fully}& OsteoHRNet & VGG-19 \cite{chen2019fully}& OsteoHRNet & VGG-19 \cite{chen2019fully}& OsteoHRNet\\
\arrayrulecolor{blue}\hline
      \end{tabular}}
  \caption{Grad-CAM visualizations generated against KL grade 4 test images using Chen \textit{et al.} \cite{chen2019fully} and OsteoHRNet.} \label{fig:GradCam4}
\end{figure*}

\subsection{Training Details}
The entire code is developed using Pytorch \cite{Pytorch} framework, and all the experiments have been conducted on a 12GB Tesla K40c GPU. Furthermore, the training of all the experimental models was optimized using stochastic gradient descent (SGD) for 30 epochs with an initial learning rate of 5e-4. Additionally, owing to the GPU capacity, the batch size was set to 24.
\subsection{Competing Methods}
In \cite{chen2019fully}, the authors proposed to utilize the pre-trained VGG-19 \cite{Simonyan15} network with a novel ordinal loss function. Yong \textit{et al.} \cite{Yong} proposed to utilize the DenseNet-161 \cite{DenseNet} with the ordinal regression module (ORM). We have compared the OsteoHRNet against the results obtained  by the best-published studies mentioned above for
a robust comparison.
\subsection{Evaluation Metrics}
In this study, we have utilized the following three evaluation metrics to analyze and compare the performance of our proposed model : (a) Multi-class accuracy, (b) Quadratic Weighted Cohen's Kappa coefficient (QWK), and (c) Mean Absolute Error (MAE). Traditionally, multi-class accuracy is defined as the average number of outcomes matching the ground truth across all the classes. Accuracy for five classes with N instances is formulated as below
\begin{equation}
    Accuracy = \frac{1}{N} \sum_{i=1}^{5} \sum_{x:g(x)=i} F(g(x) = \hat{g}(x)),
\end{equation}
where, ${F}$ is a function which returns 1 if the prediction is correct and 0 otherwise.

MAE is the mean of the absolute error of the individual prediction over all the input instances. The error in the prediction value is determined by the difference between the predicted and the true value for that given instance. MAE for five classes with N instances can be expressed as below 
\begin{equation}
    MAE = \frac{\sum_{i=1}^{N} abs({y}_{i} - {\hat{y}}_{i})}{N},    
\end{equation}
where, $y_{i}$ \&  $\hat{y}_{i}$ are the true and the predicted grade, respectively.

A weighted Cohen Kappa is a metric that accounts for the similarity between predictions and the actual values. The Kappa coefficient is a chance-adjusted index of agreement measuring the reliability of inter-annotator for qualitative prediction. The Quadratic Weighted Kappa (QWK) is evaluated using a predefined table of weights which measures the extent of non-alignment between the two raters. The greater the disagreement, the greater the weight.
\begin{equation}
    \kappa = 1- \frac{\sum_{p,\hat{p}} \textit{w}_{p,\hat{p}} O_{p,\hat{p}}}{\sum_{p,\hat{p}} \textit{w}_{p,\hat{p}} E_{p,\hat{p}}},
\end{equation}
$O$ is the contingency matrix for $K$ classes such that $O_{p,\hat{p}}$ denotes the count of ${\hat{p}}$ grade images predicted as $p$. The weight, $w$, is defined as
\begin{equation}
    \textit{w}_{p\hat{p}} = \frac{({p-\hat{p}})^{2}}{({1-K})^{2}}.
\end{equation}
Next, $E$ is calculated as the normalized product between the predicted grade's and original grade's histogram vector. Of the three metrics, accuracy and QWK are positive in nature while MAE is negative in nature.

\section{Results}\label{Results}
\subsection{Comparison against State-of-the-Art Methods}
It can be observed from Table \ref{table:result} that the proposed method has outperformed the existing best-published works \cite{chen2019fully} \cite{Yong} in terms of classification accuracy, MAE, and QWK. It should be mentioned that Yong \textit{et al.} \cite{Yong} reported the macro accuracy \footnote{Macro accuracy: 88.09\%} and contingency matrix of their best model. For a fair comparison, equivalent to the above, we have reported their results in multi-class accuracy of 70.23\%. Whereas Chen \textit{et al.} \cite{chen2019fully} has reported the best multi-class accuracy of 69.69\%. OsteoHRNet has reported a maximum multi-class accuracy of 71.74\%, multi-class average accuracy of 70.52\%, MAE of 0.311, and QWK of 0.869 which is a significant improvement over \cite{Yong}, \cite{chen2019fully}. Fig. \ref{fig:ConfusionMatrix} represents the confusion matrix obtained by using the proposed and existing methods \cite{chen2019fully}, \cite{Yong} which when fed with 1656 test images. 
\begin{table}[htbp] 
  \centering
  \caption{Quantitative comparison against the existing methods in terms of multi-class accuracy, MAE, and QWK.}
  \label{table:result}
    \resizebox{0.48\textwidth}{!}{\begin{tabular}{||c c c c||} 
 \hline
 Method & Accuracy & MAE & QWK \\ [0.5ex] 
 \hline\hline

VGG 19 - Ordinal \cite{chen2019fully} & 69.69 \% & 0.344 & 0.8460\\ [0.5ex]
 \hline
DenseNet 161 - ORM \cite{Yong} & 70.23 \% & 0.330 & 0.8609\\ [0.5ex]
 \hline
\textbf{OsteoHRNET} & \textbf{71.74 }\% & \textbf{0.311} & \textbf{0.8690} \\ [0.5ex]
 \hline
 \end{tabular}}
\end{table}%

Furthermore, we have employed the Gradient-weighted Class Activation Maps (Grad CAM) \cite{GRADCam} visualization technique to demonstrate the superiority of the proposed OsteoHRNet. It also helps in showcasing the most relevant regions the network has learned to focus on in the X-ray images. Figs. \ref{fig:GradCam0}, \ref{fig:GradCam1}, \ref{fig:GradCam2}, \ref{fig:GradCam3}, and \ref{fig:GradCam4} shows the qualitative comparison of the proposed model against the existing methods in terms of Grad-CAM visualization. It can be observed that the proposed OsteoHRNet considers both features and the area between the knee joints for an efficient severity classification  (\textit{denoted by the darker colors up the scales}). Moreover, it can be said that the decision-making of OsteoHRNet aligns in accordance with the actual real-world medical criterion of KL grade classification. The proposed model has efficiently learned the prominent features such as joint-space narrowing, osteophytes formations, and bone deformity, thus predicting the most relevant
radiological KL grading. This validates the enriched and superior results obtained by the proposed OsteoHRNet model.

\begin{figure}[t]
    \centering
    \resizebox{0.47\textwidth}{!}{
    \setlength{\tabcolsep}{0pt}
    \begin{tabular}{cc}
        \includegraphics[trim=1.2cm 0.3cm 1.2cm 0.3cm,clip]{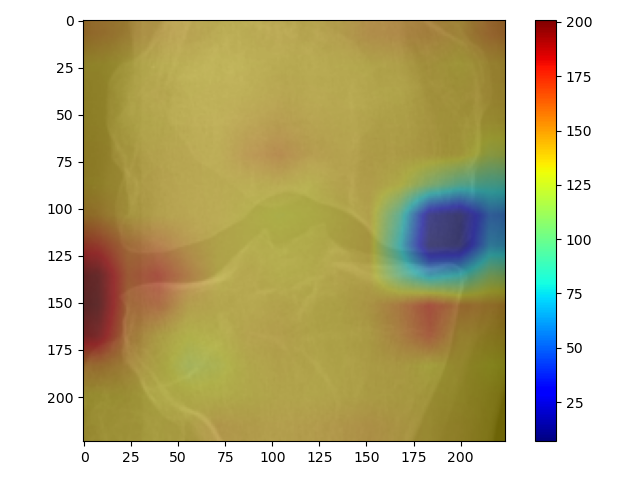} & \includegraphics[trim=1.2cm 0.3cm 1.2cm 0.3cm,clip]{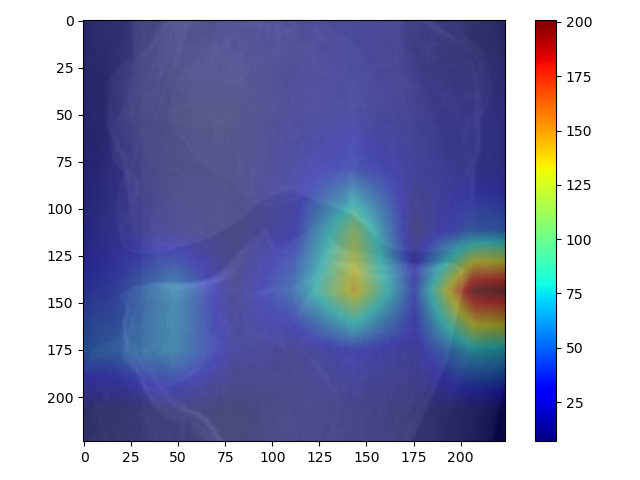}  \\
    \end{tabular}
    }
    \caption{Grad-CAM visualization for the incorrect classification by Chen \textit{et al.} \cite{chen2019fully} (VGG-19; \textbf{\textit{left}}) and proposed OsteoHRNet (\textit{\textbf{right}}) for grade 2 radiograph.}
    \label{fig:Incorrect}
\end{figure}

\begin{figure}[t]
    \centering
    \resizebox{0.47\textwidth}{!}{
    \setlength{\tabcolsep}{0pt}
    \begin{tabular}{cc}
        \includegraphics[trim=1.2cm 0.3cm 1.2cm 0.3cm,clip]{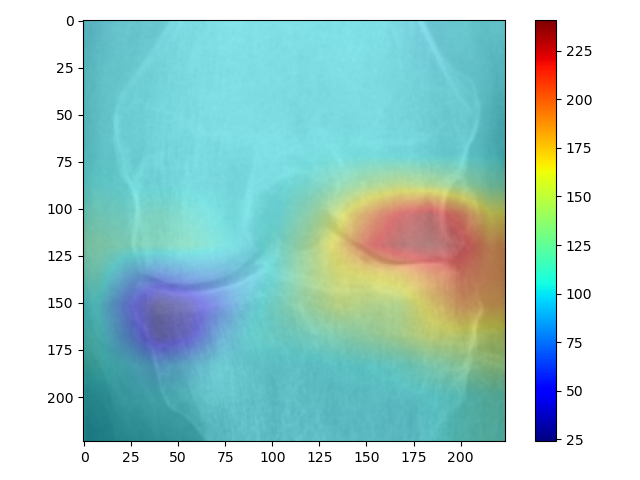} & \includegraphics[trim=1.2cm 0.3cm 1.2cm 0.3cm,clip]{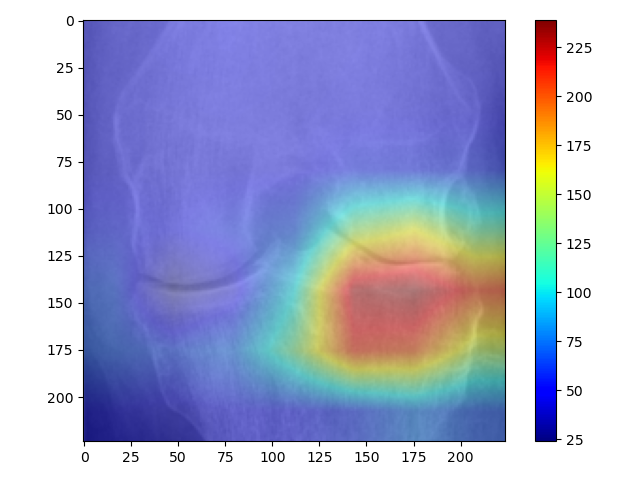}  \\
    \end{tabular}
    }
    \caption{Grad-CAM visualization for the incorrect classification by Chen \textit{et al.} \cite{chen2019fully} (VGG-19; \textbf{\textit{left}}) and proposed OsteoHRNet (\textbf{\textit{right}}) for grade 4 radiograph.}
    \label{fig:Incorrect_2}
\end{figure}

\begin{figure*} [hbtp] 
  \centering
  \setlength{\tabcolsep}{1pt}
  
  \resizebox{\textwidth}{!}{\begin{tabular}{!{\color{blue}\vrule} c c c c c!{\color{blue}\vrule}} 
 \arrayrulecolor{blue}\hline \textbf{\texttt{Grade 0}} & \textbf{\texttt{Grade 1}} & \textbf{\texttt{Grade 2}} & \textbf{\texttt{Grade 3}} & \textbf{\texttt{Grade 4}}  \\
     
       \arrayrulecolor{blue}\hline 
        \includegraphics[width=1in]{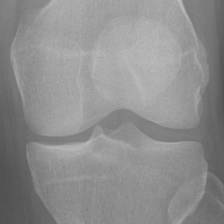} &
      \includegraphics[width=1in]{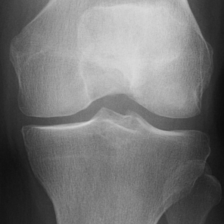} &
      \includegraphics[width=1in]{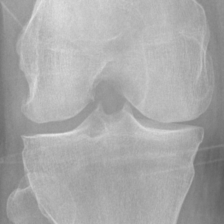} &
      \includegraphics[width=1in]{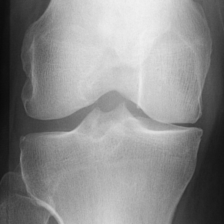} &
      \includegraphics[width=1in]{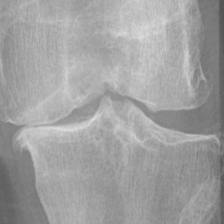} \\
      \multicolumn{5}{|c|}{\textbf{\texttt{Knee X-Ray}}}\\
      \arrayrulecolor{blue}\hline
      
       \includegraphics[width=1.35in]{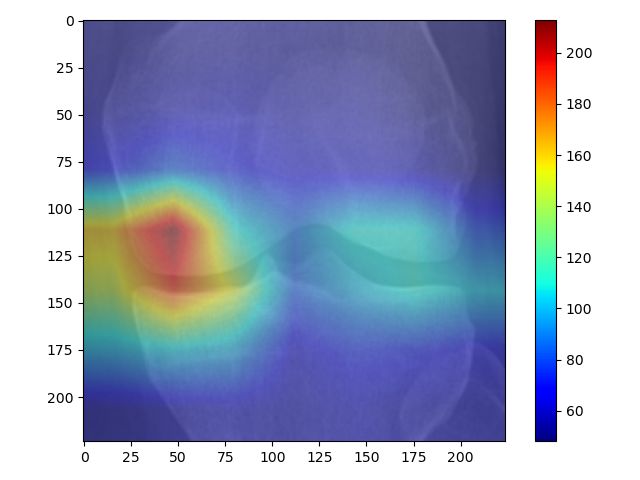} &
      \includegraphics[width=1.35in]{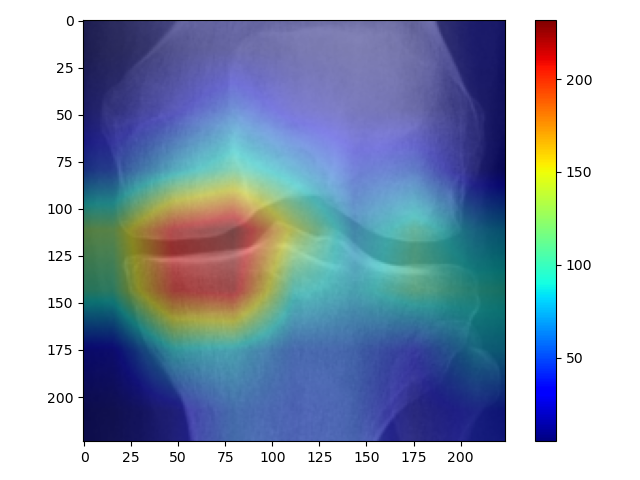} &
      \includegraphics[width=1.35in]{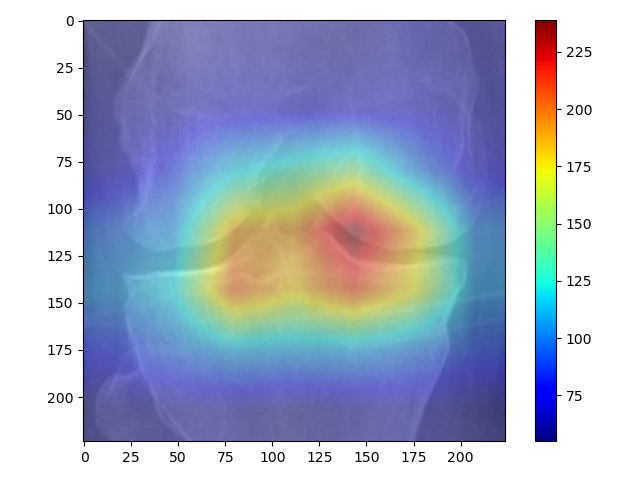} &
      \includegraphics[width=1.35in]{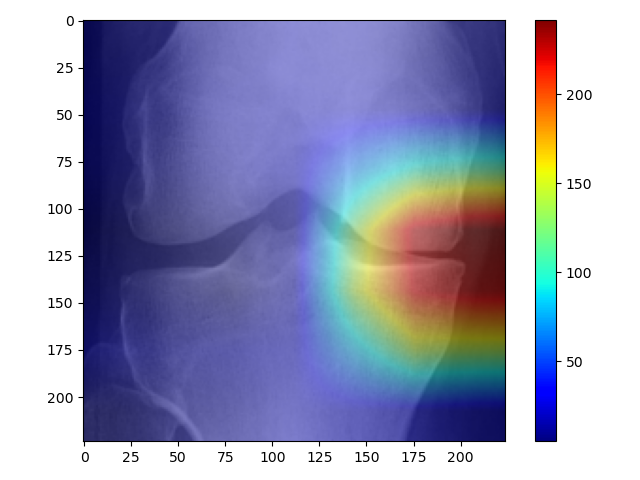} &
      \includegraphics[width=1.35in]{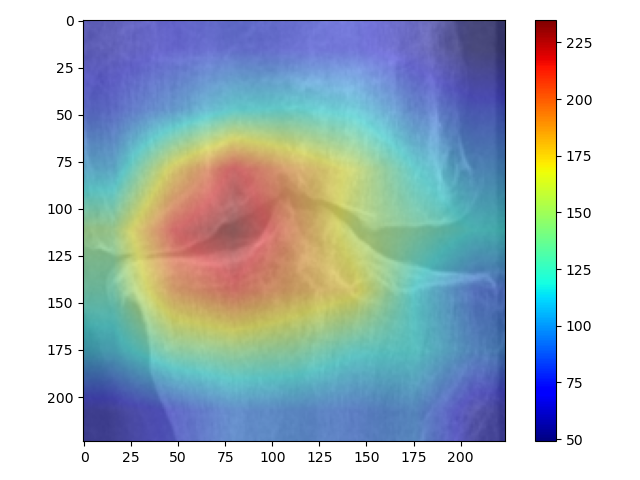} \\
      \multicolumn{5}{|c|}{\textbf{\texttt{HRNet (CE)}}}
      \\
      \arrayrulecolor{blue}\hline
      
       \includegraphics[width=1.35in]{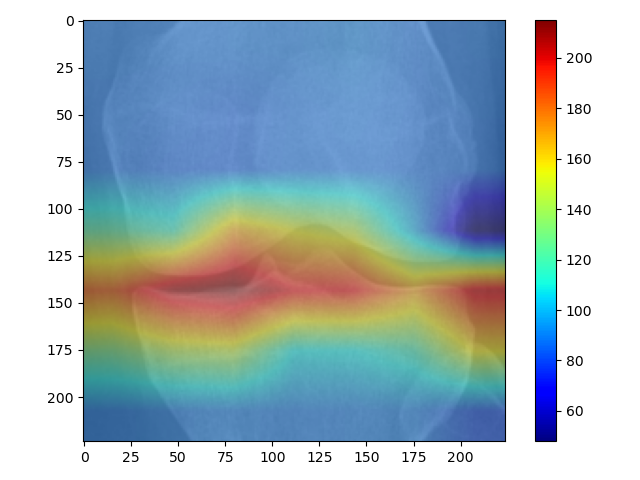} &
      \includegraphics[width=1.35in]{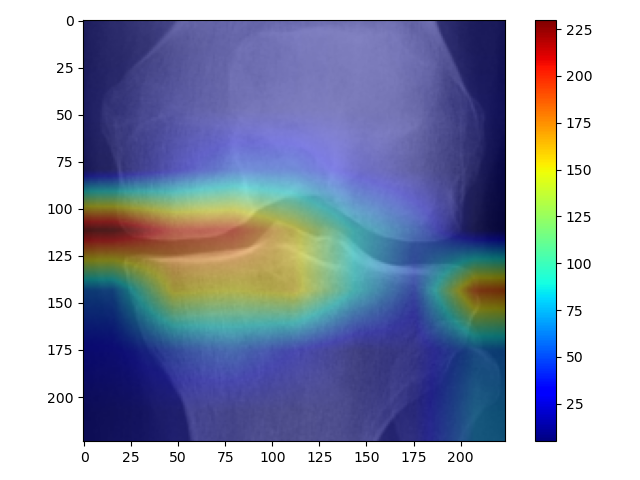} &
      \includegraphics[width=1.35in]{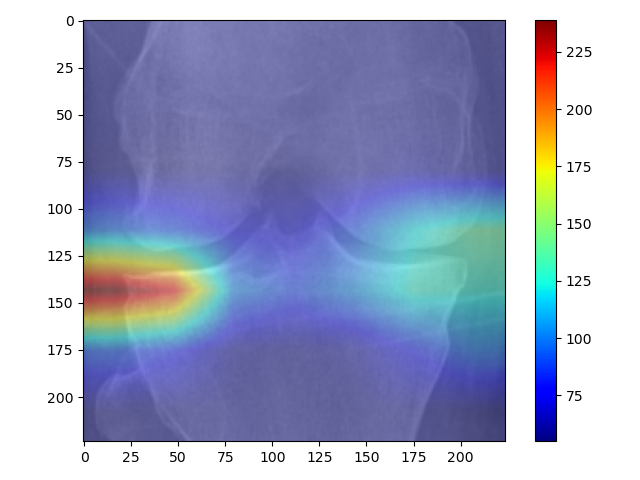} &
      \includegraphics[width=1.35in]{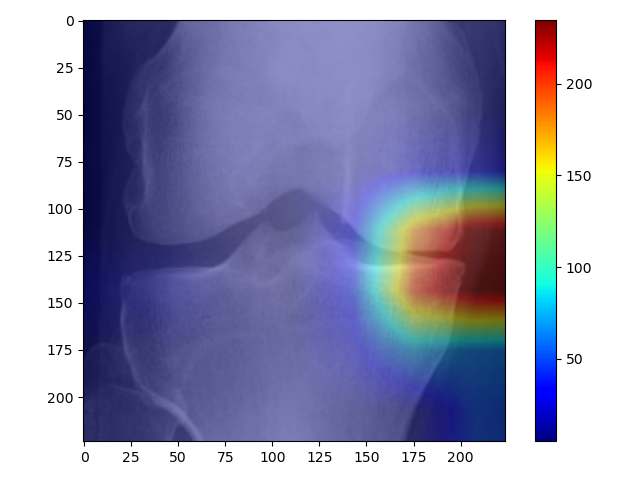} &
      \includegraphics[width=1.35in]{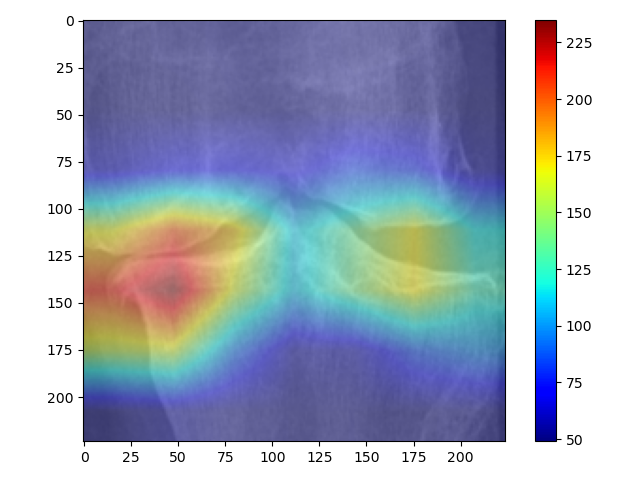} \\
      \multicolumn{5}{|c|}{\textbf{\texttt{HRNet (OL)}}}\\
      \arrayrulecolor{blue}\hline
      
       \includegraphics[width=1.35in]{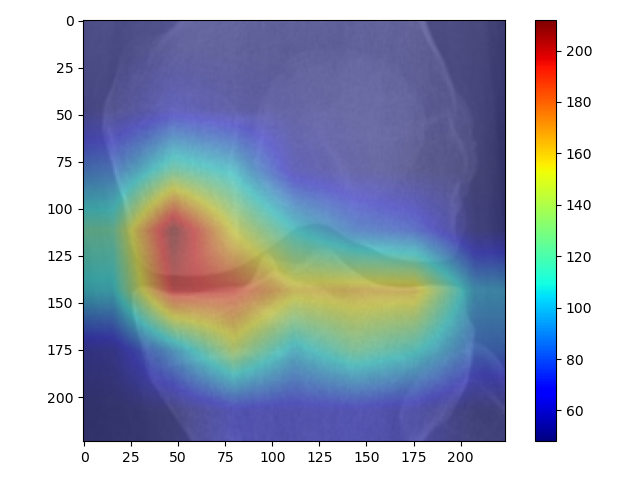} &
       \includegraphics[width=1.35in]{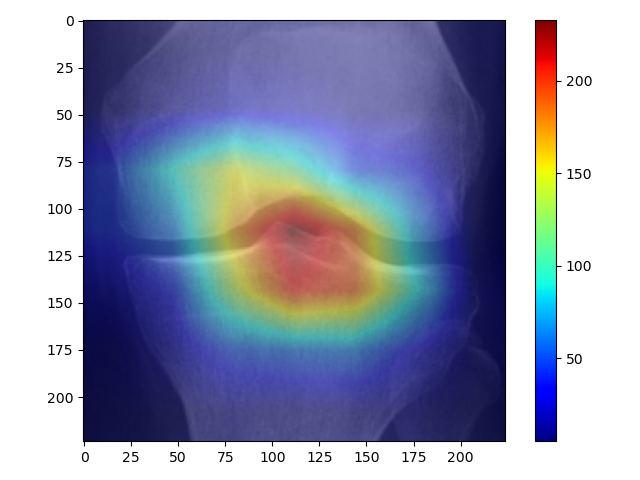} &
       \includegraphics[width=1.35in]{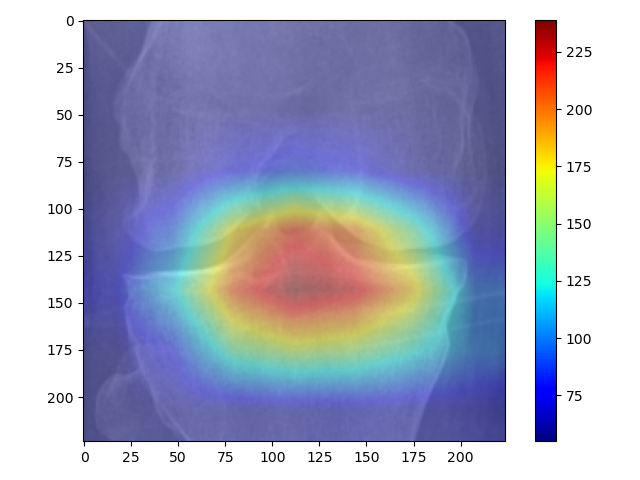} &
       \includegraphics[width=1.35in]{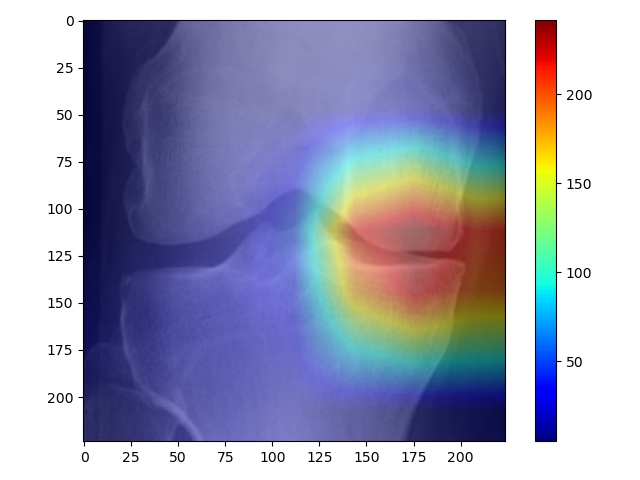} &
       \includegraphics[width=1.35in]{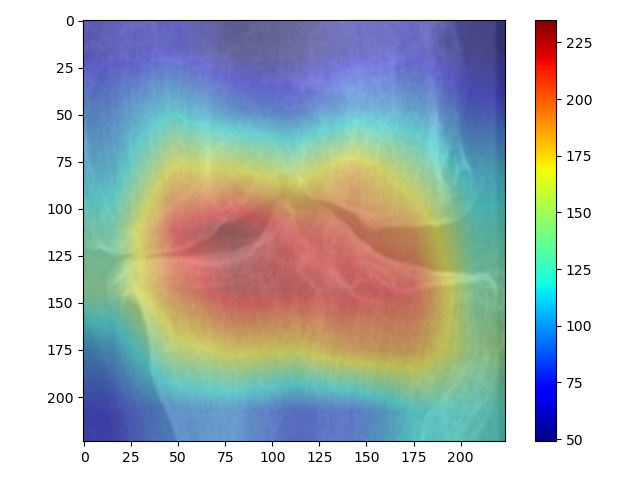} \\
      \multicolumn{5}{|c|}{\textbf{\texttt{HRNet + CBAM (CE)}}}\\ 
       \arrayrulecolor{blue}\hline
       
       \includegraphics[width=1.35in]{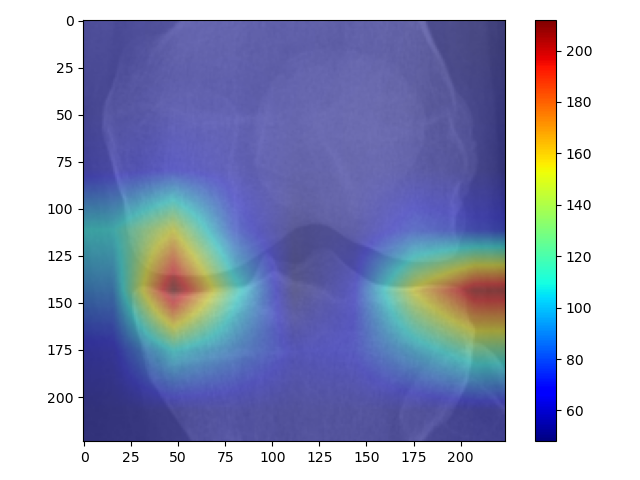} &
       \includegraphics[width=1.35in]{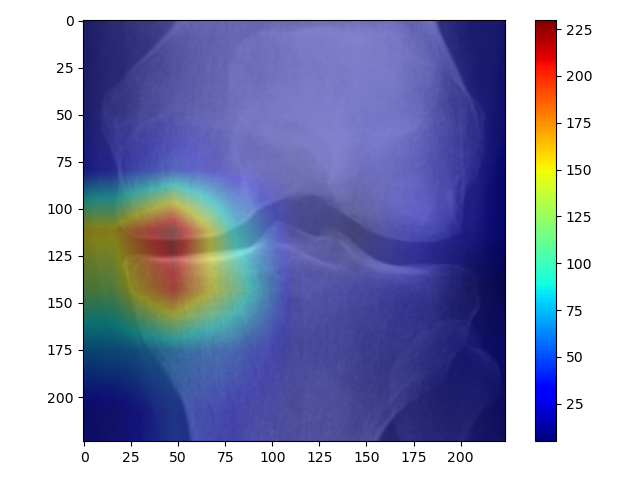} &
       \includegraphics[width=1.35in]{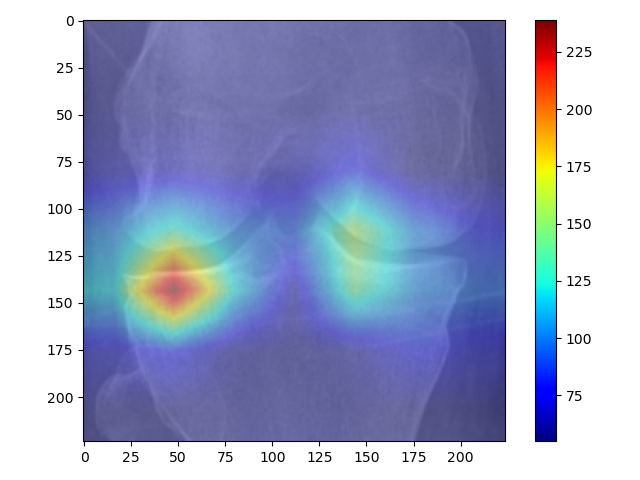} &
       \includegraphics[width=1.35in]{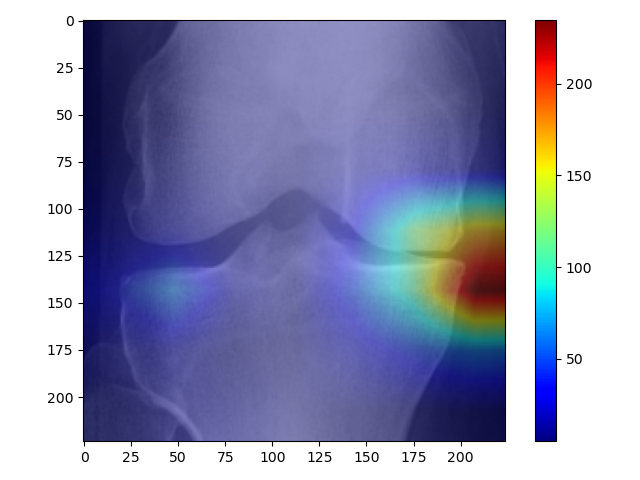} &
       \includegraphics[width=1.35in]{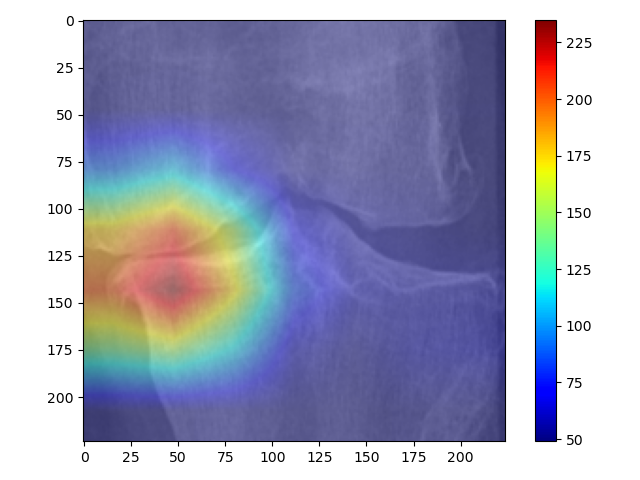} \\
       \multicolumn{5}{|c|}{\textbf{\texttt{HRNet + CBAM (OL)}}}
       \\
       \arrayrulecolor{blue}\hline
       
       \includegraphics[width=1.35in]{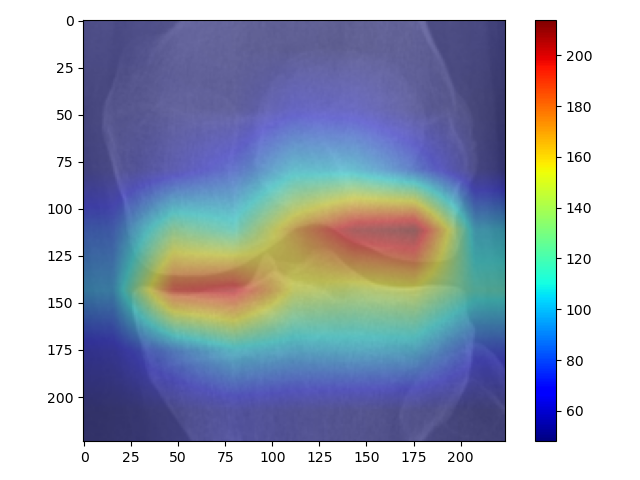} &
       \includegraphics[width=1.35in]{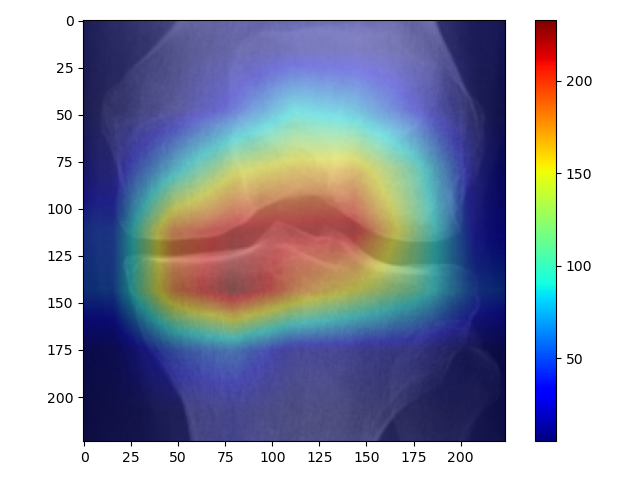} &
       \includegraphics[width=1.35in]{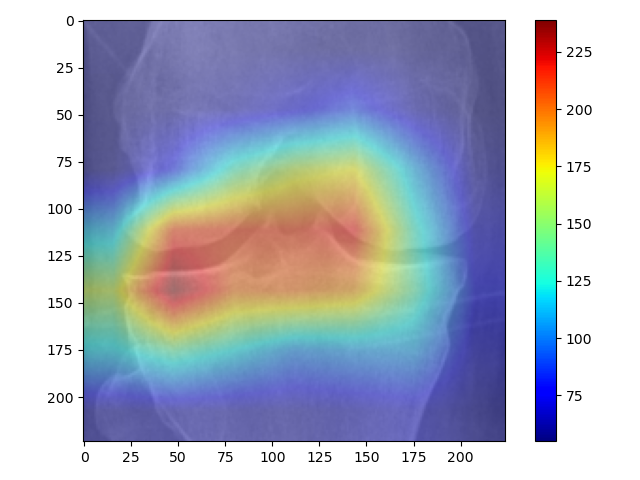} &
       \includegraphics[width=1.35in]{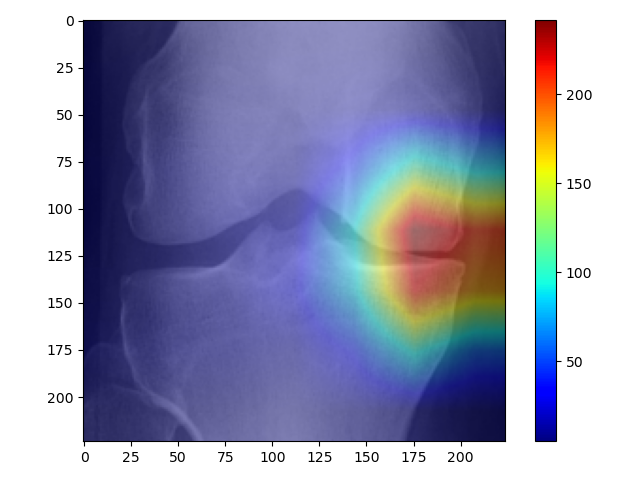} &
       \includegraphics[width=1.35in]{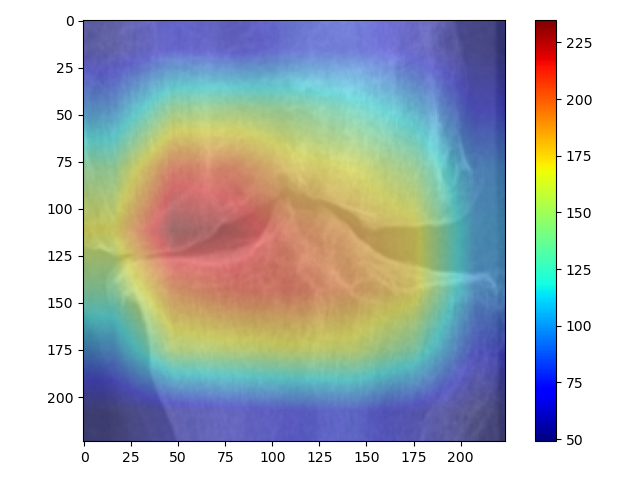} \\
       \multicolumn{5}{|c|}{\textbf{\texttt{OsteoHRNet (CE)}}}\\
       \arrayrulecolor{blue}\hline
       
       \includegraphics[width=1.35in]{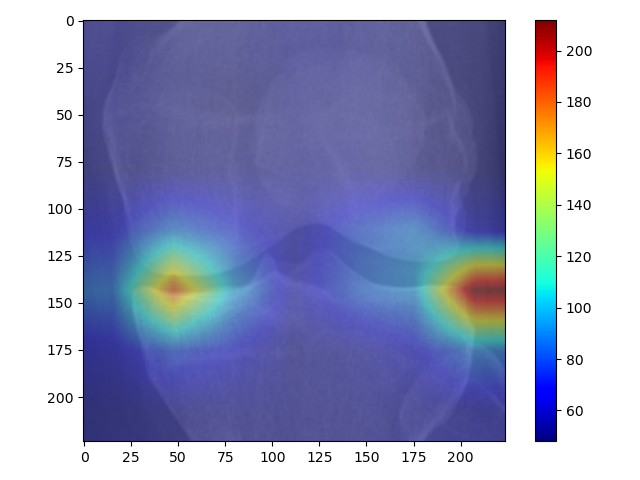} &
       \includegraphics[width=1.35in]{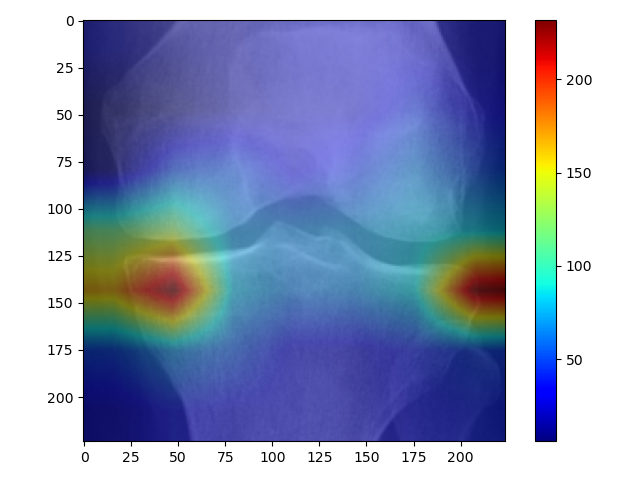} &
       \includegraphics[width=1.35in]{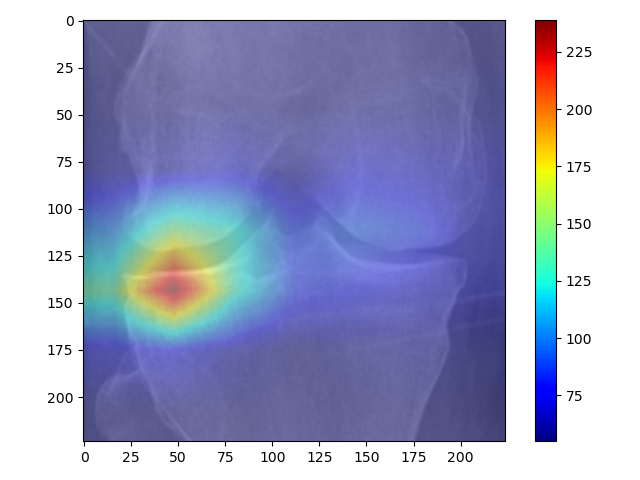} &
       \includegraphics[width=1.35in]{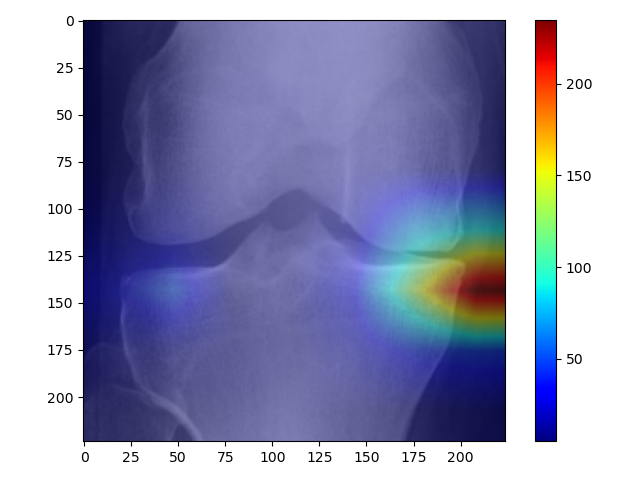} &
       \includegraphics[width=1.35in]{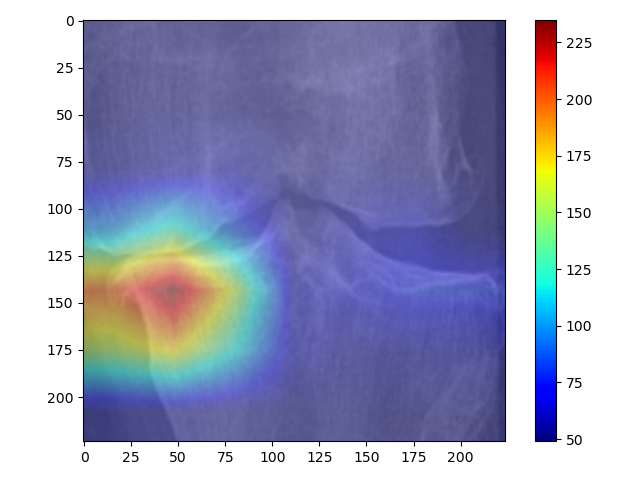} \\
       \multicolumn{5}{|c|}{\textbf{\texttt{OsteoHRNet (OL)}}}\\
       
      \arrayrulecolor{blue}\hline
  \end{tabular}}
    \caption{Grad-CAM visualizations for the ablation study. CE stands for Cross-entropy  and OL stands for Ordinal Loss} \label{table:AblationCAM}
\end{figure*}

\begin{figure*}[t]
    \centering
    \resizebox{\textwidth}{!}{
    \includegraphics[scale = 1]{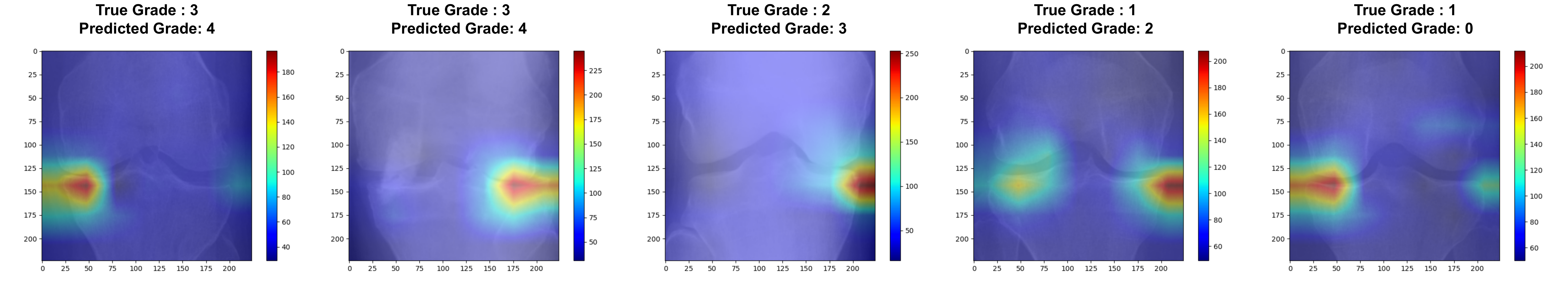}}
    \caption{Grad-CAM visualizations for the incorrectly classified radiographs ontained by using OsteoHRNet.}
    \label{fig:Incorrect_3}
\end{figure*}

\section{Discussion} \label{Discussion}
It is evident from Fig.\ref{fig:ConfusionMatrix} that the OsteoHRNet has outperformed the previous works \cite{chen2019fully},\cite{Yong}, significantly. It should be mentioned that the OsteoHRNet classifies the higher grade X-rays very accurately while reducing the misclassification between far away grades. In comparison to existing methods, there has been a significant increase in correct classifications for grade 2. Furthermore, the nearby misclassifications between higher grades (grade 2-grade 3, grade 3-grade 4) are minimum for the proposed method, which needs to be acknowledged. Also, by way of analysis using obtained Grad-CAM visualization of such incorrect classifications, it can be observed that OsteoHRNet is trying to locate joint space narrowing and osteophytes in accordance with the medical characteristics. At the same time, VGG-19 \cite{chen2019fully} is confused and focuses on the entire knee, giving importance to irrelevant features for KL grade classification, as seen in Figure \ref{fig:Incorrect}, \ref{fig:Incorrect_2}. 

Owing to its superior network learning, our model is extremely relevant to the medical setting of KL grade classification. Furthermore, the Grad-CAM visualization of our model can be extended for the use of the medical practitioner to provide confidence in the findings. 
However, our study has some limitations, and certain radiographs could not be correctly classified due to the lack of rich features in the radiographs. Fig. \ref{fig:Incorrect_3} shows nearby grade misclassifications, which to a great extent is unavoidable.  But, there is high inter and
intraobserver variability (correlation coefficient  = 0.83) for manual knee KL grading \cite{KLLimitation}. Thus, our proposed fully automated KL grading method can be extended in clinical settings for getting reliable and reproducible OA grading.  

\section{Ablation Study} \label{Ablation}
\begin{table}[htbp] 
  \centering
  \caption{Effects of different network modules \& cost function}
  \label{table:Ablation}
    \resizebox{0.48\textwidth}{!}{\begin{tabular}{||c| c c c c||} 
 \hline
 Architecture & \multicolumn{2}{|c}{Cross Entropy}  & \multicolumn{2}{c||}{Ordinal Loss}    \\  [0.75ex]
\cline{2-3} \cline{4-5} 
        & Accuracy  & MAE & Accuracy  & MAE \\ [0.75ex]
 \hline\hline

\textbf{\texttt{HRNet}} & 64.10 \% & 0.460 & 65.00 \% & 0.440\\  
\hline
\textbf{\texttt{HRNet + CBAM}} & 65.30 \% & 0.423 & 66.70 \% & 0.392\\  \hline
\textbf{\texttt{OsteoHRNet}} & 69.90 \% & 0.373 & \textbf{71.74} \% & \textbf{0.311} \\ 
 \hline
 \end{tabular}}
\end{table}

This section presents an ablation study to demonstrate the contributions made by each sub-module of the proposed OsteoHRNet. For this, we have performed the following baselines:
\begin{enumerate}
    \item \textbf{\texttt{HRNet}}: Original HRNet trained by utilizing the adopted dataset.
    
    \item \textbf{\texttt{HRNet + CBAM}}: Original HRNet followed by the CBAM module trained using the adopted dataset.
    
    \item \textbf{\texttt{OsteoHRNet}}: Original HRNet followed by the CBAM module trained using the adopted dataset. Further, during training, we have employed the data augmentation techniques to enhance the performance of the proposed model. 

\end{enumerate}

It can be observed from Table \ref{table:Ablation} that the addition of the CBAM module and data augmentation techniques have immensely improved the performance compared to its curtailed baseline. The CBAM module might have adaptively learned the relevant features from the HRNet. Such features may have contributed more towards an efficient classification compared to the features learned by the original HRNet \cite{HRNet}, VGG-19 \cite{Simonyan15}, or DenseNet161 \cite{DenseNet}.

Fig. \ref{table:AblationCAM} demonstrates the Grad-CAM visualizations for our ablation study. It can be observed that the proposed OsteoHRNet has learned the robust features progressively on each component of our proposed network. Thus, it is verified that each component of our network contributes to the final knee OA KL grade prediction. 

\section{Conclusion}\label{Conclusions}
This paper proposes a novel OsteoHRNet by
adopting the HRNet as the backbone and integrating the CBAM
module for an improved knee OA severity prediction results
from plain radiographs. The proposed network was able
to perform exceptionally well and attain significant improvements
over the previously proposed methods owing to the HRNet’s
capability to maintain high-resolution features throughout the network and its ability to capture reliable spatial features. The intermediate extracted features were significantly refined with the help of the attention mechanism; therefore, the radiographs with a similarity between classes and variations within classes could be distinguished better. Moreover, we have employed the Grad-CAM visualizations to validate that the model has learned the most relevant spatial features in the radiographs. In the future, we will work on the entire OAI multi-modal data and consider all the cohorts in our study.

\bibliographystyle{IEEEtran}

\bibliography{references}

\end{document}